\documentclass[a4paper]{article}

\long\def\drop#1{}

\usepackage{a4wide}
\usepackage[english]{babel}
\usepackage{listings}
\usepackage{amsmath,amsthm}
\usepackage{amsfonts}
\usepackage{amssymb}
\usepackage{url}

\newcommand{\eps}{\varepsilon}
\let\e\eps

\def\div{\mathop{\mathrm{div}}}
 \newcommand{\dd}{d}
\DeclareMathOperator{\supp}{supp}
\DeclareMathOperator*{\esssup}{ess\,sup}
\DeclareMathOperator*{\essinf}{ess\,inf}
\DeclareMathOperator\grad{grad}
\newcommand{\pref}[1]{(\ref{#1})}
\def\R{{\mathbb R}}
\def\N{{\mathbb N}}

\def\Dmax{D_{\mathrm{max}}}
\def\Emax{E_{\mathrm{max}}}
\def\Emin{E_{\mathrm{min}}}
\def\lebmeas{{\mathcal{L}}}

\let\downto\downarrow
\let\weakto\rightharpoonup
\def\Lebesgue{\mathcal L}
\def\RM{RM}

\newtheorem{Theorem}{Theorem}[section]
\newtheorem{Lemma}[Theorem]{Lemma}
\newtheorem{Proposition}[Theorem]{Proposition}
\newtheorem{Corollary}[Theorem]{Corollary}

\newtheorem{Definition}[Theorem]{Definition}

\def\theremark{\thesection.\arabic{Theorem}.}
\newenvironment{remark}%
    {\par\medbreak\refstepcounter{Theorem}%
         {\noindent\bf Remark~\theremark\ }}%
    {\par\medbreak}
\def\Xint#1{\mathchoice
   {\XXint\displaystyle\textstyle{#1}}%
   {\XXint\textstyle\scriptstyle{#1}}%
   {\XXint\scriptstyle\scriptscriptstyle{#1}}%
   {\XXint\scriptscriptstyle\scriptscriptstyle{#1}}%
   \!\int}
\def\XXint#1#2#3{{\setbox0=\hbox{$#1{#2#3}{\int}$}
     \vcenter{\hbox{$#2#3$}}\kern-.5\wd0}}
\def\dashint{\Xint-}

\begin{document}
\title{Well-posedness of a parabolic moving-boundary problem in the setting of Wasserstein gradient flows}
\author{Jacobus W. Portegies\thanks{Presently at the Courant Institute of Mathematical Sciences, New York University, jim@cims.nyu.edu. The work was done while the author was at the Department of Mathematics and Computer Science, Technische Universiteit Eindhoven.} 
\and
Mark A. Peletier{\thanks{Department of Mathematics and Computer Science
  and Institute for Complex Molecular Systems,
 Technische Universiteit Eindhoven, m.a.peletier@tue.nl}}}

\maketitle

\begin{abstract}
We develop a gradient-flow framework based on the Wasserstein metric for a parabolic moving-boundary problem that models crystal dissolution and precipitation. In doing so we derive a new weak formulation for this moving-boundary problem and we show that this formulation is well-posed. In addition, we develop a new uniqueness technique based on the framework of gradient flows with respect to the Wasserstein metric. With this uniqueness technique, the Wasserstein framework becomes a complete well-posedness setting for this parabolic moving-boundary problem.
\end{abstract}

\section{Introduction}

\subsection{Wasserstein formulations of parabolic PDEs}

Since the seminal work of Otto~\cite{JordanKinderlehrerOtto98,Otto01} a large number of diffusive, parabolic partial differential equations has been cast in the form of a Wasserstein gradient flow. Besides linear and nonlinear diffusion equations~\cite{Agueh05}, also convection-diffusion equations~\cite{JordanKinderlehrerOtto98}, non-local equations~\cite{CarrilloMcCannVillani06} and higher-order parabolic equations~\cite{Otto98a,GiacomelliOtto01,Glasner03} have been written as gradient flows with respect to the Wasserstein metric. Another extension is to abstract gradient flows in general metric spaces~\cite{AmbrosioGigliSavare05}.
While in many of the cases mentioned above the functional that plays the role of energy was already known, as a Lyapunov functional, the combination with the Wasserstein metric is a recent development and sheds new light on the problem. From a modelling point of view, the additional structure provided by the gradient flow has the effect of characterizing the system: once both the energy and the dissipation metric have been chosen, these choices fully determine the system, and leave no further room for variation. In this way the energy and the metric together clearly show the modelling choices that lie at the basis of the equations. The gradient-flow structure also suggests canonical time-discrete approximations, yields estimates, and provides a context in which approximations can be constructed which preserve these desirable properties.

\medskip
In this paper we extend the framework to a simple moving-boundary problem. We introduce a new weak formulation for this problem, and we exploit the Wasserstein framework to prove well-posedness of this formulation, including a contraction result in the Wasserstein metric.

We hope that this paper will be interesting to two communities. To the reader who is familiar with the weak-solution approach to moving-boundary problems, we show how the Wasserstein gradient-flow context provides a thermodynamically meaningful and mathematically sound framework in which existence, uniqueness and stability results can be derived. The thermodynamic setup provides additional insight into the behaviour of solutions.
For the reader who is familiar with the Wasserstein gradient-flow formulation of parabolic evolution equations, we show how a class of moving-boundary problems can be included, and how the Wasserstein framework provides a convenient tool for proving well-posedness.

\subsection{Crystal dissolution and precipitation}

The problem that we consider in this paper arises in the modelling of crystal dissolution and precipitation~\cite{VandeFliertVanderHout00,VanNoordenPop08}, and it represents a larger class of reactive solid-fluid interactions in which mass crosses the interface that separates the solid and fluid phases (see also~\cite{Kuiken84,SudirhamVanDammeVanderVegt06}). The model defines two species, water and solute, which may occur in mixed form (in the fluid phase) or as pure phases (as in the solid phase, which consists only of crystalline solute). At the solid-fluid interface the solid crystal may precipitate or dissolve, thus moving the interface in one direction or another. The rate of dissolution or precipitation is given by a function of the local solute concentration in the fluid.

We describe the concentrations of the two species by their volume fractions $\rho$ (for water) and $1-\rho$ (for solute). Note that this choice deviates from what is usual: the function $\rho$ is the \emph{water} volume fraction, not the solute concentration. The reason for this choice is that the problem takes a particularly simple form in terms of this variable: the solid crystal phase then corresponds to $\{x: \rho(t,x)=0\}$, and the moving interface corresponds to boundary of the fluid region $\Omega(t) = \{x:\rho(t,x)>0\}$.

The water is assumed to evolve by diffusion within the fluid region $\Omega(t)$; the boundary of this region itself evolves in time, with normal velocity $v_n$, and the boundary condition for $\rho$ is determined by the requirement that no water traverses the boundary. These assumptions lead to the equations
\begin{subequations}
\label{pb:mainNd}
\begin{alignat}2
\label{pb:mainNd:DE}
&\partial_t\rho = \Delta \rho &\qquad & \Omega(t),\\
&\frac{\partial\rho}{\partial n} = - \rho v_n && \partial \Omega(t).
  \label{pb:mainNd:Stefan}
\end{alignat}
Here $n$ and $v_n$ are the outward normal and the normal velocity of the boundary $\partial \Omega$; $n$ points from the fluid into the solid phase. For simplicity we disregard the possibility of non-moving boundaries of $\Omega$, such as rigid walls. Note that $\rho$ will typically have a non-zero limiting value at the inside of $\partial \Omega$, and we consider it to be zero outside of $\Omega$ (in the solid). Therefore $\rho$ is discontinuous at $\partial \Omega$; the value of $\rho$ appearing in~\pref{pb:mainNd:Stefan} is the interior value.

The evolution of the boundary is given by a \emph{kinetic} evolution law,
\begin{equation}
v_n = f(\rho) \qquad \partial \Omega(t),
  \label{pb:mainNd:kinetic}
\end{equation}
\end{subequations}
in which the rate function $f$ characterizes the precipitation and dissolution processes. In the context of a two-phase, water-solute system, the water volume fraction $\rho$ also characterizes the solute volume fraction $1-\rho$, and therefore we can consider the dissolution/precipitation rate $f$ a function of $\rho$. Equations (\ref{pb:mainNd:DE}-\ref{pb:mainNd:kinetic}) form a closed system for $\Omega$ and $\rho$.

Problem~\pref{pb:mainNd} is related to the so-called Stefan problem with kinetic undercooling~(see e.g.~\cite{Visintin96} or the introduction of \cite{EvansKing00a} for an overview). Both systems have a conserved quantity, the volume fraction $\rho$ in~\pref{pb:mainNd} and the enthalpy in the Stefan problem; in both cases the interfacial velocity is prescribed explicitly. The two problems represent different cases, however, corresponding to different relative rates of bulk diffusion and interfacial reaction. In the Stefan problem, reaction is fast and the heat supply slow, leading to a continuous temperature and a rate $v_n$ that depends on the local heat flux. In the problem~\pref{pb:mainNd} the diffusion and the reaction have similar rates, and the discontinuity in $\rho$ (or in the chemical potential, to be precise) drives the interfacial movement.

\subsection{A Lyapunov functional}

Typical rate functions $f$  are such that high solute values lead to precipitation, and low values to dissolution. We formalize this property by assuming that there exists a threshold volume fraction $\alpha\in(0,1)$ such that
\[
f(\rho)\gtrless 0 \qquad\Longleftrightarrow \qquad \rho\gtrless\alpha.
\]
In this case the system~\pref{pb:mainNd} has (at least formally) a Lyapunov functional
\begin{equation}
\label{def:E_nd}
E(\Omega,\rho) := \int_\Omega\bigl[ \rho\log\rho + \alpha \bigr],
\end{equation}
where it is understood that $\supp\rho\subset \Omega$. Indeed,
\begin{align*}
\frac d{dt} E(\Omega(t),\rho(t))
&= \int_{\Omega(t)} (\log \rho + 1)\partial_t \rho
  + \int_{\partial\Omega(t)} (\rho\log\rho + \alpha)v_n \\
&= \int_{\Omega(t)} (\log \rho + 1)\Delta \rho
  + \int_{\partial\Omega(t)} (\rho\log\rho + \alpha)f(\rho) \\
&= -\int_{\Omega(t)} \frac{|\nabla\rho|^2} \rho
  + \int_{\partial\Omega(t)} \Bigl[(\log \rho + 1)\frac{\partial \rho}{\partial n}
          +  (\rho\log\rho + \alpha)v_n\Bigr] \\
&= -\int_{\Omega(t)} \frac{|\nabla\rho|^2} \rho
  + \int_{\partial\Omega(t)} (-\rho+\alpha)f(\rho)\\
&\leq 0.
\end{align*}

This derivation not only shows that $E$ decreases along a solution, it also provides an expression for the rate of dissipation. Using~\pref{pb:mainNd:kinetic} we write this rate as
\begin{equation}
\label{eq:energydissipation}
\frac d{dt} E(\Omega(t),\rho(t)) = -\int_{\Omega(t)} \frac{|\nabla\rho|^2} \rho
  - \int_{\partial\Omega(t)} \frac{\rho-\alpha}{f(\rho)}v_n^2.
\end{equation}
The two terms have clearly recognizable origins: the first is associated with the diffusion of the bulk water, while the second arises from the precipation and dissolution reaction on the boundary~$\partial \Omega$. In the special case that $f(\rho)/(\rho-\alpha)$ is constant, the second term on the right-hand side depends only on $v_n$. We show in this paper that for this case, and in one space dimension, the system~\pref{pb:mainNd} can be written as a gradient flow of $E$ with respect to the Wasserstein metric; this property will be the basis for the well-posedness results that we prove. (In Appendix~\ref{app:heuristic} we also give a heuristic derivation of the gradient flow property).

\medskip
As an aside, we note that the functional $E$ can be given a simple thermodynamic interpretation. We assume that heat is conducted rapidly with respect to other processes, and that the system is kept at constant temperature $\theta$ by placing it in contact with a heat bath. In this situation the Helmholtz free energy,
\[
\int_{\R^d} \psi(\rho(t,x),\varphi(t,x))\, dx,
\qquad\text{with}\qquad
\psi(\rho,\varphi) := e(\rho,\varphi)-\theta s(\rho,\varphi),
\]
should decrease in time. In this formula $e$ and $s$ are the internal energy and the entropy of the system; $\varphi$ is a phase indicator, which takes the value $0$ in the solid and $1$ in the fluid.

In view of the formula~\pref{def:E_nd} we make the following choices for $e$ and $s$:
\[
e(\rho,\varphi) = c_1\varphi
\qquad\text{and}\qquad
s(\rho,\varphi) = -\rho\log\rho + c_2 \varphi,
\]
where $c_1$ and $c_2$ are constants. The coefficient $c_1$ is the latent heat that is absorbed upon dissolution, and it should therefore be positive; $c_2$ is an entropy penalty associated with the solid crystal, or an entropy advantage associated with dissolution, and should therefore also be positive. The function $-\rho\log\rho$ is the usual Gibbs-Boltzmann entropy for freely diffusing particles.
Upon combining these choices we find that the Helmholtz free energy equals the functional $E$, where $\alpha = c_1/\theta-c_2$.



\subsection{Weak solutions}
Solutions of~\pref{pb:mainNd} may lose regularity, for instance because of changes in topology of $\Omega(t)$, and therefore a concept of weak solution is necessary. To our knowledge, a weak formulation of problem~\pref{pb:mainNd} is currently not known. Such a weak formulation should allow for lack of regularity both in $\rho$ and in $\partial \Omega(t)$, and this is where the difficulty lies. Since this aspect is central to our work, let us explain the issue in detail.

The usual way of deriving a weak formulation proceeds by multiplying~\pref{pb:mainNd:DE} with a test function $\xi\in C^\infty_c(\R\times\R^n)$ and integrating over $\R^+\times \R^n$.  We calculate
\[
\int_0^\infty \!\!\int_{\R^n} \rho_t \xi = - \int_{\Omega(0)} \rho(0,x)\xi(0,x)\, dx
- \int_0^\infty \!\!\int_{\Omega(t)} \rho \xi_t
- \int_0^\infty \!\!\int_{\partial \Omega(t)} \rho\xi v_n,
\]
and
\[
\int_0^\infty \int_{\R^n} \xi \Delta \rho =
\int_0^\infty \int_{\partial \Omega(t)}\xi\frac{\partial \rho}{\partial n}
- \int_0^\infty \int_{\Omega(t)}\nabla \xi\nabla \rho.
\]
When combining these with~\pref{pb:mainNd:DE} and~\pref{pb:mainNd:Stefan} the boundary terms cancel, and we find
\begin{equation}
\label{def:WF0}
\int_0^\infty\int_{\Omega(t)} \bigl[\rho\xi_t - \nabla \rho\nabla \xi\bigr]
= -\int_{\Omega(0)} \rho(0,x)\xi(0,x)\, dx
\qquad \text{for all }\xi\in C^\infty_c(\R^+\times\R^n).
\end{equation}

Note that in the derivation of~\pref{def:WF0} we did not yet use the kinetic condition~\pref{pb:mainNd:kinetic}. Therefore the condition~\pref{def:WF0} applies for any given evolution of $\Omega(t)$, or put differently, the condition~\pref{def:WF0} does not yet fix the evolution of $\Omega(t)$. One of the aims of this paper is to derive a weak formulation which captures the evolution of both $\rho$ and $\Omega$, and for which existence and uniqueness results can be proved---at least in one dimension.

\subsection{The one-dimensional problem}
\label{sec:intro_1d}

We now turn to the specific problem of this paper. We study a one-dimensional version of problem~\pref{pb:mainNd}, and we restrict ourselves to solutions whose support $\Omega(t)$ is a single bounded interval $[L(t),R(t)]$. In addition we choose the specific rate function~$f(\rho) = (\rho-\alpha)/\beta$ (see Section~\ref{sec:discussion} for a discussion of this choice). Under these conditions we derive below a weak formulation which encodes both the mass-conserving transport problem~(\ref{pb:mainNd:DE}-\ref{pb:mainNd:Stefan}) and the domain evolution law~\pref{pb:mainNd:kinetic}.

In this one-dimensional version of~\pref{pb:mainNd} we seek a triplet $(\rho,L,R)$ satisfying
\begin{subequations}
\label{eq:heatmovbound}
\begin{alignat}2
&\rho_t = \rho_{xx}, & &\text{at each }(t,x) \text{ with } t>0, x\in[L(t),R(t)],\\
&\rho(0,x) = \rho_0(x), & &x \in [L_0,R_0],\\
&\rho_x(t,L(t)) = -\rho(t,L(t))L'(t),&\qquad& t>0,\\
&\rho_x(t,R(t)) = - \rho(t,R(t))R'(t), && t>0,\\
&L'(t) = \frac{\alpha - \rho(t,L(t))}{\beta}, & & t>0.
\label{eq:heatmovbound:L'} \\
&R'(t) = -\frac{\alpha - \rho(t,R(t))}{\beta}, && t>0.
\label{eq:heatmovbound:R'}
\end{alignat}
\end{subequations}
Here $\alpha$ and $\beta$ are two strictly positive constants, and subscripts denote differentiation. 

Starting from~\pref{def:WF0}, which in one dimension becomes
\[
\int_0^\infty \int_{L(t)}^{R(t)} [\rho \xi_t - \rho_x\xi_x] =
-  \int_{L(0)}^{R(0)} \rho(0,x)\xi(0,x)\, dx,
\]
we perform another partial integration on the second term to find
\begin{equation}
\label{eq:weakform-inprogress}
\int_0^\infty \int_{L(t)}^{R(t)} [\rho \xi_t + \rho\xi_{xx}]
- \int_0^\infty [\rho(t,R(t))\xi_x(t,R(t)) -\rho(t,L(t))\xi_x(t,L(t))]
=
-  \int_{L(0)}^{R(0)} \rho(0,x)\xi(0,x)\, dx.
\end{equation}
The trace of $\rho$ on the boundary $x=R(t)$ can be rewritten using the boundary condition~\pref{eq:heatmovbound:R'},
\begin{equation}
\label{eq:rewrittenboundaryvalue}
\rho(t,R(t))\xi_x(t,R(t)) = \beta R'(t)\xi_x(t,R(t)) + \alpha\xi_x(t,R(t)).
\end{equation}
and the middle term in this expression can again be rewritten by using the identity
\begin{equation}
\label{eq:totderivxi}\frac{\dd}{\dd t} \xi(t,R(t)) = \xi_t(t,R(t)) +
 \xi_x(t,R(t)) R'(t).
\end{equation}

Applying these remarks to~\pref{eq:weakform-inprogress} results in the following definition. By $\lebmeas(\R^+)$ we mean the set of Lebesgue measurable functions on $\R^+$; $E$ is the energy defined above, i.e.
\[
E(\rho,L,R) := \int_L^R \rho\log\rho + \alpha(R-L).
\]

\begin{Definition}\label{de:weak} (weak solutions)\\
We call $(\rho,L,R) \in L^\infty(\mathbb{R}^+;L^1(\mathbb{R}))\times
\lebmeas(\mathbb{R}^+)\times \lebmeas(\mathbb{R}^+)$ a \emph{weak
solution} to problem (\ref{eq:heatmovbound}) if

\begin{enumerate}
\item \label{cond:weaksol:1}
$\rho \geq 0$, and for all $t \geq 0$, $L(t)<R(t)$ and $\supp \rho(t,\cdot) \subset
[L(t),R(t)]$;
\item \label{cond:weaksol:2}
$\Emax := \esssup_{t>0} E(\rho(t),L(t),R(t)) < \infty$;
\item \label{cond:weaksol:3}
for all $\xi\in C_c^\infty(\mathbb{R}^2)$
\begin{multline}
\label{eq:weakform1}- \int_{L(0)}^{R(0)} \rho_0(x) \xi(0,x) \dd x -
\beta \xi(0,R(0)) - \beta \xi(0,L(0)) \\ - \int_0^\infty
\int_{L(t)}^{R(t)} \rho(t,x) \xi_t(t,x) \dd x dt   - \beta\int_0^\infty
\xi_t(t,R(t))\dd t - \beta\int_0^\infty \xi_t(t,L(t)) \dd t =
\\ =  \int_0^{\infty} \int_{L(t)}^{R(t)}
\rho(t,x)\xi_{xx}(t,x)\dd x dt  - \alpha \int_0^\infty \xi_x
(t,R(t)) \dd t + \alpha \int_0^\infty \xi_x(t,L(t))\dd t.
\end{multline}
\end{enumerate}

\end{Definition}

\begin{remark}
Note that the weak form~\pref{eq:weakform1} now only requires integrability of $\rho$; no derivatives or trace values are used. Similarly, the domain limits $L$ and $R$ are not required to be continuous in time. Of course, higher-regularity results may give us much more regularity than this; for instance, the properties of the heat equation imply that the function $\rho$ is smooth inside the domain, and Lemma~\ref{lemma:contraction} shows that solutions automatically have continuous boundaries. The main point here is however that the formulation itself does not require additional regularity.
\end{remark}

\begin{remark}
The second condition in Definition~\ref{de:weak} is new to the discussion. For a classical solution of a gradient-flow system the energy decreases with time; for such a solution the global upper bound on the energy reduces to a finite-energy condition on the initial datum. Any generalization of the solution concept should preserve the monotonicity of the energy, and indeed the energy also decreases along solutions in the sense of Definition~\ref{de:weak}.

The energy bound plays a role in proving uniqueness of weak solutions. The superlinear control on $\rho$ that follows from this bound provides continuity of the function $t\mapsto \rho(t)$ in the topology $\sigma(L^1,L^\infty)$ (see Lemma~\ref{lemma:contraction}), and this additional regularity is just enough for the proof of contraction.
\end{remark}

\begin{remark}
Since we do not explicitly assume any regularity in time other than measurability of weak solutions, the proof of their uniqueness will be different from uniqueness proofs found in for instance~\cite{AmbrosioGigliSavare05}. Indeed, the latter results consider uniqueness in the class of functions that are absolutely continuous in the time variable. Moreover, uniqueness is proven for solutions of the so-called Evolution Variational Inequality~\cite[(4.0.13)]{AmbrosioGigliSavare05}, and not directly for the differential equation~(\ref{eq:weakform1}). On the other hand, in~\cite{AmbrosioGigliSavare05} uniqueness results are obtained without the second assumption and with a slightly more general initial condition. Therefore, we cannot state that one result is stronger than the other.
\end{remark}

\begin{remark}
Weak solutions that are sufficiently regular also are solutions of~\pref{pb:mainNd} in the classical sense.
\end{remark}

\drop{
Before moving to the questions of existence, uniqueness, and stability of weak solutions, we first show the consistency of this definition, by proving that weak solutions with additional regularity also satisfy the strong formulation~\pref{pb:mainNd}.

\begin{Lemma}
Let $\rho \in C^{1,2}(\mathbb{R}^+\times \mathbb{R},\mathbb{R}^+)$,
and $L,R \in C^1(\mathbb{R},\mathbb{R})$ be a weak solution according to Definition~\pref{de:weak}. Then $(\rho,L,R)$ is a classical solution of~\pref{eq:heatmovbound}.
\end{Lemma}

\begin{proof}
The main issue is to derive from a weak solution the two interface conditions~\pref{pb:mainNd:Stefan} and~\pref{pb:mainNd:kinetic}. By applying partial integration to~\pref{eq:weakform1} and using~\pref{eq:totderivxi} we obtain
\begin{align}
\nonumber \int_{L_0}^{R_0} (-\rho_0(x)+\rho(0,x) ) \xi(0,x)&\dd
x\\+\nonumber \int_0^\infty \int_{L(t)}^{R(t)} (\rho_t(t,x) &-
\rho_{xx}(t,x) ) \xi(t,x)  \dd x \dd t = \\
\nonumber &\int_0^\infty(- \rho(t,R(t))R'(t) -
\rho_x(t,R(t)) ) \xi(t,R(t)) \dd t \\
\nonumber & + \int_0^\infty(\rho(t,L(t)) L'(t) +
\rho_x(t,L(t)) ) \xi(t,L(t)) \dd t \\
\nonumber & + \int_0^\infty (-\beta R'(t)+
\rho(t,R(t))-\alpha)\xi_x(t,R(t)) \dd t\\
\nonumber &  + \int_0^\infty (- \beta L'(t) - \rho(t,L(t))+\alpha)
\xi_x(t,L(t))\dd t.
\end{align}
By making various choices for $\xi$ we find that $(\rho,L,R)$ is a classical solution of
(\ref{eq:heatmovbound}).
\end{proof}
}

\subsection{Main results and layout of the paper}

We start the next part of this paper by introducing some basic concepts and notation, most importantly the Wasserstein distance and some of its properties. We then turn to the  main result of this paper:
\begin{Theorem}
\begin{itemize}
\item
Let $L_0<R_0$, $\rho_0\in L^1(\R)$, $\rho_0\geq0$, and $\supp\rho_0\subset[L_0,R_0]$. Then there exists a weak solution $(\rho,L,R)$ with initial datum $(\rho_0,L_0,R_0)$.
\item The solution operator is a contraction in the Wasserstein metric.
\end{itemize}
\end{Theorem}

The contraction is proved as Theorem~\ref{th:contraction} in Section~\ref{sec:contraction}. As an intermediate result we derive an evolutionary variational inequality~\pref{EVI} from which the contraction follows readily. The main ingredients for the proof of this evolutionary inequality are the Kantorovich duality formulation of the Wasserstein distance and a reformulation of the weak-solution definition formula~\pref{eq:weakform1}.  In Section~\ref{sec:existence} we use the well-known method of time discretization to prove the existence of weak solutions (Theorem~\ref{th:existence}). We have chosen to include this proof for the specific case rather than to refer to a more general theory in order to be self-contained and accessible. However, in Section~\ref{sec:AGS}, we will place the problem in the context of the general theory by Ambrosio, Gigli, and Savar\'e of gradient flows in metric spaces~\cite{AmbrosioGigliSavare05}. Doing so, we easily obtain some extra results, such as an energy identity and regularity results.

In Appendix~\ref{app:heuristic} we provide a heuristic discussion that illustrates in a different way how the system of this paper can be viewed as a Wasserstein gradient flow. This discussion also motivates the specific choice $f(s) = (s-\alpha)/\beta$.

\subsection{Discussion}
\label{sec:discussion}

Some of our results are not new: existence and uniqueness
for classical solutions of a very similar one-dimensional problem has been proved by Van de Fliert and Van der Hout~\cite{VandeFliertVanderHout00}, and general methods for Stefan-like problems can also be applied~\cite{Friedman64}. The restriction to one dimension also allows for a formulation in terms of the variable $U(x) := \int_{-\infty}^ x\rho$, which is very similar to a classical Stefan problem~\cite{VazquezPersonal}. Interestingly, the convenient expression of the Wasserstein distance in terms of the inverse of this function $U$ shows that this approach is actually related to that of this paper. On the other hand, other results are new, such as the weak formulation and its well-posedness, and the Wasserstein contraction property of the solution operator.

In addition to proving new results, however, we also aim to demonstrate that this gradient-flow framework is a natural setting for this problem. This claim of `naturalness' becomes apparent in a number of different places. To start with, the energy $E$ has a meaningful thermodynamic interpretation. In fact one can also start with the energy $E$ and the appropriate dissipation penalization (given by~\pref{def:g} in Appendix~\ref{app:heuristic}) and derive the evolution equations~\pref{pb:mainNd} from these two choices. In addition, since no other gradient-flow frameworks are known for this problem, the mere existence of this structure is remarkable.

The second indication of `naturalness' arises in the fact that we prove existence, uniqueness, and stability of weak solutions all within the context of Wasserstein gradient flows. Only requiring very weak regularity of potential solutions, the necessary compactness properties follow from intrinsic properties of the energy $E$ and the Wasserstein distance.

Finally, a third indication of naturalness can be found in the
derivation of the contraction property from the definition of a weak solution. Although slightly obscured by technicalities, the contraction theorem springs from a convenient combination of the test-function behaviour~\pref{eq:h'} with the Kantorovich duality characterization~\pref{eq:dualW_2}. Together these yield the central inequality~\pref{EVI}. In subsequent work we have developed this concept further~\cite{NatilePeletierSavare10TR}.

\medskip

However, we have to accept two major limitations in order to make this scheme work: the specific choice of $f$ and the single space dimension. These are fundamentally linked to both the existence of the weak solution and the use of the Wasserstein distance.

\textbf{The rate function $f$.}
The choice $f(\rho)=(\rho-\alpha)/\beta$ initially arises as the only possibility in the derivation of the weak formulation: the manipulation of the terms in~(\ref{eq:rewrittenboundaryvalue}--\ref{eq:totderivxi}) requires exactly this form in order to be successful. Interestingly, the same choice is special in the dissipation balance~\pref{eq:energydissipation}: for any other choice of $f$, the trace of $\rho$ on $\partial\Omega$ appears in this equation.  Whether the double appearance of this form of $f$ is connected at some level is an intriguing question.

\textbf{The single space dimension. }
%
In this paper, movement of the boundaries $L$ and $R$ is penalized by adding two Dirac delta functions (with weight $\beta$) to the function $\rho$. For such an augmented probability distribution the Wasserstein distance appropriately penalizes both movement in $\rho$ and in the boundary positions.

In higher dimensions the interfacial area will vary with time, and therefore simply weighting the interface with `$\beta$ times the $(n-1)$-dimensional Hausdorff measure' is not possible: the Wasserstein distance between two measures of unequal mass is undefined. Modifications such as varying the weight $\beta$ such that the total mass is constant, or transporting only the common part of the mass~\cite{CaffarelliMcCann06TR}, appear to have unphysical effects.

The central question is how to convert the dissipation tensor~\pref{def:g} into a metric distance. The same dissipation tensor appears in the (formal) description of motion by mean curvature as a gradient flow of the area functional; in the context of this problem some steps have been made. Luckhaus introduced a time-discretisation of the Stefan problem with Gibbs-Thomson law~\cite{Luckhaus91,LuckhausSturzenhecker95}; R\"oger later modified this time-discrete problem by replacing global minimization by a form of local minimization~\cite{Roeger04}. Almgren, Taylor and Wang used a similar approximation in the context of crystal growth~\cite{AlmgrenTaylorWang93}. These time-discrete minimization problems have the formal structure of the common backward-Euler approximations of gradient flows~\cite{AmbrosioGigliSavare05}, and they appear to reduce, in the limit of small step size, to~\pref{def:g}. However, the rigorous definition of an associated metric distance still presents difficulties.

In addition, the derivation of the weak formulation does not generalize from one to higher dimensions. The higher-dimensional equivalent of \pref{eq:totderivxi} reads
\[
\frac d{dt}\int_{\partial\Omega(t)} \xi\, ds
= \int_{\partial\Omega(t)} \left[ \xi_t + \nabla \xi\cdot n v_n - \xi H\cdot n v_n\right]\, ds,
\]
where $H$ is the total curvature of the boundary. The additional curvature term prevents us from applying the one-dimensional arguments to higher dimensions, at least in the current form.

\medskip

In conclusion, in this paper we use the Wasserstein metric to represent not only the dissipation in the diffusing bulk but also the dissipation at the reacting boundary. This appears to be unique to one dimension and to a specific rate function. However, the formal structure of a gradient flow of $E$ with respect to a dissipation metric of Wasserstein type---see Appendix~\ref{app:heuristic}---is generally valid. This leaves some hope for generalizations of the scheme to higher dimensions as well and it will be an interesting challenge to place these generalizations on a rigorous footing. For now, we will carefully work out the scheme in the one-dimensional case.



\section{Energy and metric}
\label{sec:notation}

In this paper a `measure' is always a non-negative Borel measure, and $\RM(\R)$ is the space of Radon measures, i.e. measures $\mu$ such that $\mu(\R)<\infty$. Where necessary, the one-dimensional Lebesgue measure is denoted $\Lebesgue$.

\subsection{The state space $G$}
The unknown in the definition of a weak solution above is a triplet $(\rho,L,R)$. Throughout this paper it will be useful to bundle the three components into a single object,
which is an element of the space
\[
G = \left\{ \mu \in RM(\R) :\;\mu = \rho \mathcal L + \beta \delta_L + \beta \delta_R,\  \rho\geq0, \ \int \rho = 1, \ \supp \rho \subset [L,R]\right\},
\]
i.e. if $\mu\in G$, then for $\phi\in C(\R)$
\[
\int \phi \, d\mu = \int_\R \rho(x)\phi(x)\, dx + \beta\phi(L)+\beta\phi(R).
\]
We will write this identification as $\mu=(\rho,L,R)$. Note that with this notation, applied both to the solution $\rho$ and the initial datum $\rho_0$, the equation~\pref{eq:weakform1} that defines a weak solution can also be written as
\begin{multline}
\label{eq:weakform1a}
- \int \xi(0,y)\, d\mu_0(y)
- \int_0^\infty \int\xi_t(t,y)\,d\mu(t;y)dt =
\\ =  \int_0^{\infty} \int_{L(t)}^{R(t)}
\rho(t,y)\xi_{yy}(t,y)\,\dd y dt  - \alpha \int_0^\infty \xi_y
(t,R(t)) \,dt + \alpha \int_0^\infty \xi_y(t,L(t))\,dt.
\end{multline}

\subsection{Energy}
As above the energy $E : G \to \R$ is defined by
\begin{equation}
\label{eq:InternalEnergy}
E(\mu) := \int_\mathbb{R} \rho \log \rho +
 \alpha (R - L).
\end{equation}
where $\mu=(\rho,L,R)$.

\begin{Lemma} (Energy bounds)
\label{P:lbintenergy} For every $\mu \in G$,
\[
E(\mu) \geq \Emin := \log \alpha + 1,
\]
and if $E(\mu)\leq M$, then there exists $M' = M'(M,\alpha)$ such that
\begin{equation}
\label{bound:energytoterms}
\int \rho\log\rho \leq M \qquad\text{and}\qquad
R-L \leq M'.
\end{equation}
\end{Lemma}
\begin{proof}
Using Jensen's inequality we calculate
\begin{align}
\nonumber E(\mu) &= (R-L) \int_L^R \frac{\rho(x) \log \rho(x)}{R-L} \dd x
+ \alpha(R-L)\\
&\geq \log\left(\frac{1}{R-L}\right)+ \alpha (R-L) \geq
\log \alpha + 1.
\label{bound:lower_calc}
\end{align}
where the last
inequality is a property of the function $x\mapsto \log(1/x) +
\alpha x$.
To prove~\pref{bound:energytoterms} we note that~\pref{bound:lower_calc} implies that
\[
-\log (R-L)+ \alpha(R-L) \leq M
\]
and since the function $x\mapsto -\log x + \alpha x$ is unbounded as $x\to\infty$, the result follows.
\end{proof}

\begin{Corollary} \label{C:bndsep}
For every weak solution $\mu=(\rho,L,R)$, we have $\sup_{t>0}(R(t)-L(t)) <\infty$.
\end{Corollary}

\subsection{The Wasserstein distance}
We now turn to the Wasserstein metric.
Let $\mu_0$ and $\mu_1$ be measures on $\mathbb{R}$ satistfying $\mu_0(\R)=\mu_1(\R)$. We say that a measure $\gamma$ has
$\mu_0$ and $\mu_1$ as its marginals, if one of the following
equivalent conditions holds:

\begin{itemize}
\item For all Borel sets $A \subset \mathbb{R}$,
\[
\gamma[A \times \mathbb{R}] = \mu_0[A] \text{ and } \gamma[
\mathbb{R}\times A] = \mu_1[A].\\
\]
\item For $(\varphi, \psi) \in L_{\mu_0}^1(\mathbb{R})\times
L_{\mu_1}^1(\mathbb{R})$,
\[
\int_{\mathbb{R}\times \mathbb{R}} [ \varphi(x) + \psi(y)] \,\dd
\gamma(x,y) = \int_{\mathbb{R}} \varphi(x) \,\dd \mu_0(x) +
\int_{\mathbb{R}} \psi(y) \,\dd \mu_1(y).
\]
\end{itemize}
The set of all measures satisfying these conditions we
call $\Gamma(\mu_0,\mu_1)$.
We say that a measurable function $T:\mathbb{R}\rightarrow\mathbb{R}$ \emph{pushes $\mu_0$ forward to $\mu_1$}, and we write $\mu_1=T_\#\mu_0$, if
\begin{itemize}
\item $\mu_1[A]=\mu_0[T^{-1}(A)]$ for all Borel sets $A\subset \mathbb{R}$, or equivalently\\
\item for all $\varphi \in L^1(\mu_1)$,
\[
\int_{\mathbb{R}} \varphi(y) \,\dd
\mu_1(y) = \int_{\mathbb{R}} \varphi(T(x)) \,\dd \mu_0(x).
\]
\end{itemize}

If $\mu_0$ and $\mu_1$ also satistfy $\int_\R x^2 d\mu_0(x)< \infty$ and $\int_\R y^2d\mu_1(y)<\infty$, then the Wasserstein distance is
defined by
\[
W_2(\mu_0,\mu_1)^2 := \inf
\left\{\int_{\mathbb{R}^d\times\mathbb{R}^d}(x-y)^2 \dd \gamma(x,y) \quad
| \quad \gamma \in \Gamma(\mu_0,\mu_1) \right\}.
\]
The measure $\gamma$ that achieves this infimum is called the \emph{optimal transport plan}~\cite{Villani03}. A useful characterization of optimality is the following~\cite[Theorem 5.9]{Villani08}: An admissible transport plan $\gamma\in \Gamma(\mu^0,\mu^1)$ is optimal iff $\supp\gamma $ is \emph{monotonic}, i.e
\[
\forall (x_1,y_1),(x_2,y_2)\in \supp\gamma, \qquad
(x_1-x_2)(y_1-y_2)\geq 0.
\]
As a result we have the simple characterization for distances in $G$,
\begin{equation}
\label{decomp_W}
W_2^2(\mu^0,\mu^1) = \beta(L^0-L^1)^2 + W_2^2(\rho^0,\rho^1) +\beta(R^0-R^1)^2.
\end{equation}

The distance~$W_2$ also has the dual representation~\cite[Theorem~5.9]{Villani08}
\begin{equation}
\label{eq:dualW_2}
\frac12 W_2(\mu_0,\mu_1)^2 = \sup_{\varphi\in L^1(\mu_0)} \left\{
  \int_\R (\tfrac12 x^2 - \varphi(x)) \, d\mu_0(x)
      + \int_\R (\tfrac12 y^2 - \varphi^*(y))\, d\mu_1(y)
  \right\}.
\end{equation}
Here $\varphi^*$ is the convex conjugate (the Legendre transform) of $\varphi$. If the optimal transport plan~$\gamma$ can be characterized in terms of an \emph{optimal transport map}, i.e. a convex function $\varphi$ such that
\[
\int_{\R\times\R} \zeta(x,y)\, d\gamma(x,y)
= \int_{\R\times\R} \zeta(x,\varphi'(x))\, d\mu_0(x)
\qquad\text{for any $\zeta\in C_c(\R^2)$},
\]
then the supremum in~\pref{eq:dualW_2} is reached at $\varphi$.

\medskip
Equipped with the metric $W_2$ the set $G$ is a metric space. Convergence in this space is equivalent to weak-* convergence of the measures together with a uniform bound on their support (see e.g.~\cite[Theorem~7.12]{Villani03}):
\[
W_2(\mu_n,\mu) \longrightarrow 0 \quad\Longleftrightarrow\qquad
\left\{\begin{array}{l}
\int \phi(d\mu_n-d\mu) \longrightarrow 0\qquad \text{for all }\phi\in C_c(\R),\text{ and}\\
\sup_n |L_n|,|R_n|<\infty.
\end{array}\right.
\]
We shall also use
\begin{Lemma}
\label{lemma:Elsc}
The energy $E$ is lower-semicontinuous in the Wasserstein metric.
\end{Lemma}

\begin{proof}
Let $W_2(\mu_n,\mu)\to0$ as $n\to0$. By~\pref{decomp_W} this implies that each of $|L_n-L|$, $|R_n-R|$, and $W_2(\rho_n,\rho)$ vanishes. The term $R-L$ in~\pref{eq:InternalEnergy} then converges, and by the lower-semicontinuity of $\rho \mapsto \int \rho\log\rho$ (e.g.~\cite{McCann97}), we have
\[
\int \rho\log\rho\leq \liminf_{n\to\infty} \int \rho_n\log\rho_n.
\]
\end{proof}

\subsection{Regular elements of $G$}
An important way of regularizing elements of $G$ is to replace them by approximations for which $\rho$ is bounded from above and away from zero on $[L,R]$.

\begin{Definition}
An element $\mu = (\rho,L,R) \in G$ is called \emph{regular} if $\rho\in L^\infty(\R)$ and  $\essinf_{[L,R]}\rho>0$.
\end{Definition}

If one end point is regular, the Wasserstein distance can be characterized in a particularly useful way.

\begin{Lemma}
\label{lemma:char_varphi}
Let $\mu^0,\mu^1\in G$ and assume that $\mu^1$ is regular. Then there exists a convex function $\varphi\in W^{2,1}_{\mathrm{loc}}(\R)$ such that
\begin{equation}
\label{eq:bdrypoints}
\varphi_x(L^0)= L^1\qquad\text{and}\qquad \varphi_x(R^0) = R^1,
\end{equation}
and
\begin{equation}
\label{eq:W2-equality}
\tfrac12 W_2^2(\mu^0,\mu^1) = \int(\tfrac12 x^2 - \varphi(x))\, d\mu^0(x)
  + \int (\tfrac12 y^2-\varphi^*(y))\, d\mu^1(y).
\end{equation}
\end{Lemma}

The proof is given in the appendix. Using this characterization we exploit the \emph{displacement convexity}~\cite{McCann97} of $E$ to prove

\begin{Lemma}
Let $\mu^0,\mu^1\in G$ and let $\mu^1$ be regular. Then
\begin{equation}
\label{ineq:convexity}
E(\mu^1)-E(\mu^0) \geq \int_\R (1-\varphi''(x))\rho^0(x)\, dx
 + \alpha (R^1-R^0-L^1+L^0),
\end{equation}
where $\varphi$ is the function given by the previous lemma. Here $\int\varphi''\rho^0$ should be interpreted as $+ \infty$ if $\varphi''\rho^0$ is not integrable.
\end{Lemma}

\begin{proof}
The displacement convexity of $\rho\mapsto\int \rho\log\rho$ gives
\[
\int \rho^1\log\rho^1 - \int\rho^0\log\rho^0 \geq \int (1-\varphi''(x))\rho^0(x)\, dx.
\]
This follows for instance from taking the function $\psi$ in~\pref{limit:tangent-E} equal to $x\mapsto x-\varphi'(x)$. The second term on the right-hand side of~\pref{ineq:convexity} is immediate.
\end{proof}

\section{Contraction}
\label{sec:contraction}
In this section we show that weak solutions satisfy a contraction property: the Wasserstein distance between two solutions decreases in time. The fact that the energy $E$ is displacement convex suggests such a contraction property (see e.g.~\cite{Otto01,AmbrosioGigliSavare05}). The proof that we give is special in that it only requires the relatively weak solution concept of Definition~\ref{de:weak}. In other words, the contraction properties are derived from the weak form of the differential equation itself. In particular, we have assumed no regularity in time for such solutions other than measurability and in this respect the following treatment differs from a typical treatment found in~\cite{AmbrosioGigliSavare05}.

\begin{Theorem}
\label{th:contraction}
Let $\mu^0=(\rho^0,L^0,R^0)$ and $\mu^1=(\rho^1,L^1,R^1)$ be two weak solutions. Then the function
\[
t\mapsto  W_2(\mu^0(t),\mu^1(t))
\]
is non-increasing for $t\in[0,\infty)$.
\end{Theorem}

We start with some preliminary lemmas.

\begin{Lemma}
\label{lemma:contraction}
Let $\mu=(\rho,L,R)$ be a weak solution.
\begin{enumerate}
\item Then
\begin{equation}
\label{eq:equal_mass}
\int_{L(t)}^{R(t)} \rho(t,x)\, dx = \int_{L_0}^{R_0} \rho_0(x)\, dx,
\qquad \text{for almost all }t>0,
\end{equation}
and
\begin{equation}
\label{eq:const_first_moment}
\int x\, d \mu(t;x) = \int x\, d \mu(0;x), \qquad \text{for almost all } t>0.
\end{equation}
Moreover, there exists a constant $\Dmax$ such that $|R(t)|,|L(t)| \leq \Dmax$.
\item \label{lemma:contraction:h}
The function
\[
t\mapsto \mu(t)
\]
(which is defined for almost every $t>0$) has a representative which is $W_2$-continuous on $[0,\infty)$; the absolutely continuous part $t\mapsto \rho(t)$ is $\sigma(L^1(\R),L^\infty(\R))$-continuous on $[0,\infty)$. In addition $L,R\in C([0,\infty))$.
\item \label{lemma:contraction:h'}
For any $\zeta\in W_{\mathrm{loc}}^{2,1}(\R)$ with $\sup_{x\in\R}\zeta''(x)<\infty$, the function
\[
h_\zeta(t) := \int \zeta(x)\,d\mu(t;x)
\]
satisfies for every $t>0$
\begin{equation}
\label{eq:h'}
\lim_{\e\downto0} \frac{h_\zeta(t)-h_\zeta(t-\e)}{\e}
\leq \int_{L(t)}^{R(t)} \zeta''(x) \rho(t,x)\, dx +\alpha\zeta'(L(t)) - \alpha\zeta'(R(t)).
\end{equation}
\end{enumerate}
\end{Lemma}

\begin{proof}
To start, we set $\xi(t,x) := \psi(t)\zeta(x)$ in~\pref{eq:weakform1a}, with $\psi\in C_c^1((0,\infty))$ and $\zeta\in C^2_c(\R)$. Then
\begin{equation}
\label{eq:curly_braces}
-\int_0^\infty \psi'(t) \int \zeta(x)\, d\mu(t;x)dt =
\int_0^\infty \psi(t) \left\{
  \int_{L(t)}^{R(t)}\zeta''(x)\rho(t,x)\, dx
     + \alpha\zeta'(L(t)) - \alpha \zeta'(R(t))
\right\}dt.
\end{equation}
Equalities~\pref{eq:equal_mass} and~\pref{eq:const_first_moment} follow from specific choices for $\zeta$. Choosing the sequence $\zeta_\e(x) := \zeta(\e x)$ with $\zeta(0)=1$ and taking the limit $\e\to0$ yields~\pref{eq:equal_mass}; choosing the sequence $\zeta_\e(x) := \zeta(\e x)/\e$ with $\zeta(0)=0$, $\zeta'(0)=1$ yields~\pref{eq:const_first_moment}. This latter inequality, combined with the bound on $R-L$ (Corollary~\ref{C:bndsep}), gives the separate boundedness of $L$ and $R$.

For the second part, choose a representative for the function $t\mapsto\mu(t)$ (again denoted $\mu$). For any $\zeta\in C^2_c(\R)$ and the expression in curly braces in~\pref{eq:curly_braces} is bounded uniformly in $t$; therefore $h_\zeta\in W^{1,\infty}(0,\infty)$. Therefore there is a continuous representative (also denoted $h_\zeta$) such that for $t\in(0,\infty)\backslash N_\zeta$, with $N_\zeta$ a null set,
\begin{equation}
\label{eq:definition-h}
h_\zeta(t) = \int_\mathbb{R} \zeta(x) \dd \mu(t,x),
\end{equation}
and
\begin{equation}
\label{eq:h'-in-proof}
h_\zeta'(t) = \int \zeta''(x) \rho(t,x)\, dx +\alpha\zeta'(L(t)) - \alpha\zeta'(R(t)).
\end{equation}

Let $D\subset C^2_c(\R)$ be a countable dense subset of $C_c(\R)$. Then we can define a null set $N$ such that~\pref{eq:definition-h} holds for every $\zeta\in D$ and every $t\in(0,\infty)\backslash N$. We claim that if $\{t_n\}_{n\in \N}\subset (0,\infty)\setminus N$ with $t_n\to t_0$, then $\mu(t_n)$ $W_2$-converges to a limit, and that this limit is independent of the choice of sequence (but depends, of course, on $t_0$). Assuming this for the moment, we can extend $\mu$ to a $W_2$-continuous function on $(0,\infty)$ by taking limits.

To prove the claim above, first assume two time sequences both converge to $t_0>0$, such that $\mu$ converges in $W_2$ to different limits $\mu_1$ and $\mu_2$. We then choose $\zeta\in D$ such that $\int \zeta d\mu_1\not=\int\zeta d\mu_2$; this contradicts the continuity of $h_\zeta$ proved above. This shows that limits are independent of the sequence. Since the measures $\{\mu(t)\}_{t\in(0,\infty)\backslash N}$ are tight, sequences $\{\mu(t_n)\}$ are $W_2$-compact, and therefore contain a convergent subsequence; by the uniqueness proved above the whole sequence converges.

To prove the $\sigma(L^1,L^\infty)$-continuity we use a similar argument. By part~\ref{cond:weaksol:2} of Definition~\ref{de:weak} any sequence $(\rho(t_n))$ is precompact in the $\sigma(L^1,L^\infty)$ topology (see e.g.~\cite{DunfordSchwartz58,Brezis83,AmbrosioFuscoPallara00}); the uniqueness of the limit follows from the uniqueness of the $W_2$-limit. This concludes part~\ref{lemma:contraction:h} of the Lemma.

For part~\ref{lemma:contraction:h'} we fix the time $t$. Note that by~\pref{eq:h'-in-proof} for $\zeta\in C^2_c(\R)$ and fixed $\e>0$,
\begin{align}
\notag
\frac{h_\zeta(t)-h_\zeta(t-\e)}{\e} &= \dashint_{t-\e}^t \left\{
  \int_{L(t)}^{R(t)} \zeta''(x)\rho(s,x)\, dx + \alpha(\zeta'(L(s))-\zeta'(R(s)))
  \right\} ds\\
&= \int_\R \zeta''(x) \dashint_{t-\e}^t\rho(s,x)\, dsdx +
  \alpha\dashint_{t-\e}^t (\zeta'(L(s))-\zeta'(R(s)))\, ds
\label{eq:diff-q-h}
\end{align}
For general $\zeta\in W_{\mathrm{loc}}^{2,1}(\R)$ we smoothly truncate $\zeta$ to zero outside of $[-\Dmax-1,\Dmax+1]$ and define $\zeta_n := \psi_n * \zeta$, where $\{\psi_n\}$ is a sequence of mollifiers. Then $h_{\zeta_n}(t)\to h_\zeta(t)$, $h_{\zeta_n}(t-\e)\to h_\zeta(t-\e)$, $\zeta_n''\to \zeta''$ in $L^1$, and $\zeta_n'\to\zeta'$ uniformly on $\R$. Because of the boundedness of the support of $\mu$, the terms in~\pref{eq:diff-q-h} are independent of the choice of truncation.

Next take the limit $\e\downto0$. By the continuity of $L$ and $\zeta'$,
\[
\lim_{\e\downto0} \dashint_{t-\e}^t \zeta'(L(s))\, ds = \zeta'(L(t)).
\]
Since for any $m>0$ we have
\[
\int_\R \zeta''(x) \dashint_{t-\e}^t\rho(s,x)\, dsdx
\leq \int_\R \max\{\zeta''(x),-m\} \dashint_{t-\e}^t\rho(s,x)\, dsdx,
\]
and since the function $x\mapsto \max\{\zeta''(x),-m\} \in L^\infty(\R)$, it follows from the $\sigma(L^1,L^\infty)$-continuity that
\[
\lim_{\e\downto0} \int_\R \zeta''(x) \dashint_{t-\e}^t\rho(s,x)\, dsdx
  \leq \int_\R \max\{\zeta''(x),-m\} \rho(t,x)\, dx.
\]
The left-hand side is independent of $m$, and therefore by the monotone convergence theorem
\[
\lim_{\e\downto0} \int_\R \zeta''(x) \dashint_{t-\e}^t\rho(s,x)\, dsdx
  \leq \int_\R \zeta''(x) \rho(t,x)\, dx.
\]
Collecting the parts yields the assertion.
\end{proof}

\begin{Lemma}
\label{lemma:EVI}
Let $\mu^0:[0,\infty)\to G$ be a weak solution and let $\mu^1\in G$ be arbitrary. Then the function $t\mapsto W_2^2(\mu^0(t),\mu^1)$ satisfies for all $\mu^1\in G$ the
\emph{evolution variational inequality}
\begin{equation}
\frac12\frac d{dt} W_2^2(\mu^0(t),\mu^1) + E(\mu^0(t))
\leq  E(\mu^1),
\label{EVI}
\end{equation}
in distributional sense.
\end{Lemma}

\begin{proof}
We will show that $\mu^0$ satisfies
\begin{multline}
\tfrac12 W_2^2(\mu^0(t_2),\mu^1) - \tfrac12 W_2^2(\mu^0(t_1),\mu^1)
\leq (t_2-t_1)E(\mu^1) - \int_{t_1}^{t_2} E(\mu^0(s))\, ds \\
\qquad\text{for all }0<t_1<t_2\text{ and all }\mu^1\in G.
\label{wEVI}
\end{multline}

First, we show that it is sufficient to prove~\pref{wEVI} under the additional assumption that $\mu^1$ is regular. Suppose that~\pref{wEVI} has been proved for all regular~$\mu^1$. Approximate a general $\mu^1\in G$ by regular $\mu^1_n$, such that
\[
W_2(\mu^1,\mu^1_n)\to0 \qquad\text{and}\qquad
\rho^1_n\to\rho^1 \text{ in }L^1(\R).
\]
This can be done, for instance, by restricting $\rho^1$ on $[L^1,R^1]$ to the value set $[1/n,n]$, and then renormalizing the result to regain the same mass as $\rho^1$. Then for all $t>0$,
\[
W_2(\mu^0(t),\mu^1) = \lim_{n\to\infty} W_2(\mu^0(t),\mu^1_n) \qquad\text{and}\qquad
E(\mu^1)= \lim_{n\to\infty} E(\mu^1_n).
\]
Therefore we can pass to the limit in~\pref{wEVI} for $\mu^1_n$ and obtain the same inequality for $\mu^1$.

We therefore continue under the assumption that $\mu^1$ is regular. Lemma~\ref{lemma:char_varphi} provides a function $\varphi(t,\cdot)\in W^{2,1}_{\mathrm{loc}}(\R)$ that satisfies the equality~\pref{eq:W2-equality} at time $t$; by using the same function $\varphi(t,\cdot)$ in the characterization~\pref{eq:dualW_2} for time $t-\e$,  we find
\[
\tfrac12 W_2^2(\mu^0(t),\mu^1) - \tfrac12 W_2^2(\mu^0(t-\e),\mu^1)
\leq \int(\tfrac12x^2 -\varphi(t,x))(d\mu^0(t;x)-d\mu^0(t-\e;x)).
\]
Then
\begin{align}
\label{est:central}
\limsup_{\e\downto0} &\frac1{2\e} \left[W_2^2(\mu^0(t),\mu^1) - W_2^2(\mu^0(t-\e),\mu^1)\right]\notag\\
&\leq \limsup_{\e\downto0} \frac1\e \int(\tfrac12x^2 -\varphi(t,x))(d\mu^0(t;x)-d\mu^0(t-\e;x))\notag\\
&\stackrel{\pref{eq:h'}}\leq \int(1-\varphi_{xx}(t,x))\rho^0(t,x)\, dx
  + \alpha \Bigl[L^0(t) - \varphi_x(t,L^0(t)) - R^0(t) + \varphi_x(t,R^0(t))\Bigr]\notag\\
&\stackrel{\pref{eq:bdrypoints}}= \int(1-\varphi_{xx}(t,x))\rho^0(t,x)\, dx
  + \alpha \Bigl[L^0(t) - L^1 - R^0(t) + R^1\Bigr]\notag\\
&\stackrel{\pref{ineq:convexity}}\leq E(\mu^1) - E(\mu^0(t)).
\end{align}
Therefore the function
\[
f(t) := \tfrac12 W_2^2(\mu^0(t),\mu^1) + \int_0^t[E(\mu^0(s))-E(\mu^1)]\, ds
\]
is continuous and has a non-positive left derivative at every $t>0$; by an application of the next  lemma it follows that $f$ is decreasing, which is a reformulation of~\pref{wEVI}.
\end{proof}
\begin{Lemma}
\label{lemma:decreasing_function}
Let $f\in C(\R)$ satisfy
\[
\limsup_{\e\downarrow0} \frac1\e (f(t)-f(t-\e)) \leq 0,\qquad \text{for all } t\in\mathbb{R},
\]
Then $f$ is non-increasing.
\end{Lemma}

\begin{proof}
Assume, to force a contradiction, that there exist $t_1<t_2$ with
\[
\ell := \frac{f(t_2)-f(t_1)}{t_2-t_1} > 0.
\]
Define
\[
g(t) := f(t) - f(t_1) - \ell (t-t_1).
\]
The function $g$ is continuous, and therefore takes its maximum over the interval $[t_1,t_2]$ at some $t_3\in [t_1,t_2]$. Note that $g(t_1)= g(t_2) = 0$, so that we can assume that $t_3>t_1$. Since $g$ is maximal at $t_3$,
\[
\liminf_{\e\downarrow0} \frac1\e (g(t_3) - g(t_3-\e)) \geq 0.
\]
On the other hand,
\[
\limsup_{\e\downarrow0} \frac1\e (g(t_3) - g(t_3-\e))
 = -\ell + \limsup_{\e\downarrow0} \frac1\e (f(t_3) - f(t_3-\e)) \leq -\ell,
\]
which implies a contradiction.
\end{proof}

\begin{proof}[Proof of Theorem~\ref{th:contraction}]
Defining
\[
d(s,t) := \frac12 W_2^2(\mu^0(s),\mu^1(t))
\qquad\text{for } s,t>0,
\]
the proof of the theorem is concluded by showing that $d/dt \, d(t,t)\leq 0$ in the sense of distributions. This is done by an argument of~\cite[Lemma~4.3.4]{AmbrosioGigliSavare05}, which we reproduce here for completeness.

Let $\zeta\in C_c^1((0,\infty)), \zeta\geq0$, and calculate for sufficiently small $\e>0$
\begin{align*}
-\int_0^\infty d(t,t) \frac{\zeta(t)-\zeta(t-\e)}\e \, dt
&= \int_0^\infty \zeta(t)\frac{d(t,t)-d(t-\e,t-\e)}\e \, dt\\
&= \int_0^\infty \zeta(t)\frac{d(t,t)-d(t-\e,t)}\e \, dt
 + \int_0^\infty \zeta(t+\e)\frac{d(t,t+\e)-d(t,t)}\e \, dt.
\end{align*}
Now note that by~\pref{wEVI} we can deduce
\begin{align*}
-\int_0^\infty d(t,t) \frac{\zeta(t)-\zeta(t-\e)}\e \, dt
&\leq \int_0^\infty \zeta(t)\frac1\e\left( \e E(\mu^1(t)) - \int_{t-\e}^t E(\mu^0(s)) \,ds \right)  \, dt \\
 &\quad + \int_0^\infty \zeta(t+\e)\frac1\e\left( \e E(\mu^0(t)) - \int_t^{t+\e} E(\mu^1(s)) \,ds \right) \, dt.
\end{align*}
The integrands at the right hand side are bounded from above. Moreover, by the $W_2$-continuity of $\mu^{0,1}$ there exists a constant $C>0$ such that
\[
|d(t,t)| \leq C
\qquad\text{for all }t,(t-\e) \in \supp\zeta \text{ and $\e$ sufficiently small.}
\]
The dominated convergence theorem, Fatou's Lemma, and the fact that $E$ is lower semi-continuous with respect to the Wasserstein metric (Lemma~\ref{lemma:Elsc}), yield
\begin{align*}
-\int_0^\infty d(t,t) \zeta'(t)\, dt
&\leq \int_0^\infty \zeta(t) \limsup_{\e\downto0} \left(E(\mu^1(t))-\dashint_{t-\e}^t E(\mu^0(s))\,ds)\right)\, dt\\
& \quad + \int_0^\infty \zeta(t) \limsup_{\e\downto0} \left(E(\mu^0(t))-\dashint_t^{t+\e} E(\mu^1(s))\,ds\right)\, dt\\
&\leq \int_0^\infty \zeta(t) (E(\mu^1(t))-E(\mu^0(t))+E(\mu^0(t))-E(\mu^1(t)))\, dt \leq 0.
\end{align*}
\end{proof}

\section{Existence of solutions}
\label{sec:existence}

The proof of existence of a solution follows the well-known argument of time-discretization. Similar proofs in the Wasserstein context have been given in~\cite{Otto98,Otto01,Agueh05}. We provide the proof here for completeness.

\begin{Theorem}
\label{th:existence} Let $\mu_0\in G$. Then there exists a weak solution of problem
(\ref{eq:heatmovbound}) with initial datum $\mu_0$ as defined in Definition~\ref{de:weak}.
\end{Theorem}

\def\step#1#2{\par\smallskip\noindent \textbf{Step #1 (#2).}}
The main steps in the proof of Theorem~\ref{th:existence} are

\step1{Time-discrete minimization problem} Let $h>0$ and $\sigma\in G$ be fixed, and define
\begin{equation}
\label{eq:Ifunctional} I_{h,\sigma}(\mu) = \frac{1}{2h}
W_2^2(\sigma,\mu) + E(\mu),\quad \text{for }\mu \in G.
\end{equation}
\begin{Lemma}
\label{lemma:exminpb}
There exists a unique minimizer $\mu=(\rho,L,R)\in G$ of $I_{h,\sigma}$. It satisfies the Euler-Lagrange equation
\begin{equation}
\label{eq:approxEL1}
\left| \frac{1}{h}\int \zeta(x) \,\dd (\mu(x) -
\sigma(x))
 - \int_\mathbb{R} \rho(x) \zeta'' (x) \,\dd x  + \alpha
\zeta'(R) - \alpha \zeta'(L) \right|\\ \leq \frac{1}{2h} \|
\zeta'' \|_\infty W_2^2(\sigma,\mu)
\end{equation}
for every
$\zeta \in C_c^\infty(\mathbb{R})$.
\end{Lemma}

\step2{Approximate solutions} For $h>0$ fixed, and for given initial datum $\mu_0\in G$, construct the sequence $(\mu^h_i)_{i\geq0}$ by
\[
\mu^h_{i+1} \text{ minimizes }I_{h,\mu^h_i}, \qquad
\text{and}\qquad\mu^h_0 := \mu_0.
\]
Define the approximate solution $\mu^h:[0,\infty)\to G$ by
\[
\mu^h(t) := \mu^h_i \qquad \text{if}\qquad
t\in [ih,(i+1)h).
\]
\begin{Lemma}
\label{lemma:approxEL}
For all $\zeta \in C_c^\infty(\mathbb{R} \times \mathbb{R})$, and $T$
such that $\zeta(t,x)=0$ whenever $t\geq T$,
\begin{multline}
\label{eq:almostweak} \left| - \int_0^T \int
\partial_t^h \zeta(t,x) \dd \mu^h(t;x)  \,\dd t - \frac{1}{h} \int_0^h
\int \zeta(t,x) \dd \mu_0(x) \,\dd t \right. \\
\left. - \int_0^T \int_\mathbb{R} \rho^h(t,x)\zeta_{xx}(t,x)\, \dd x \dd
t
+ \alpha \int_0^T \zeta_x(t,R^h(t)) \,\dd t-\alpha \int_0^T \zeta_x(t,L^h(t)) \,\dd t \right| \\
\leq \frac{1}{2} \| \zeta_{xx} \|_\infty \sum_{i=1}^{\lceil T/h\rceil}
W_2^2(\mu_{i-1}^h, \mu_i^h),
\end{multline}
where the discrete derivative
$\partial_t^h$ is defined by
\[
\partial_t^h\zeta(t,x) := \frac{\zeta(t+h,x)-\zeta(t,x)}{h}.
\]
\end{Lemma}

\step3{Estimating the error term}
\begin{Lemma}
\label{lemma:est_error}
For all $h>0$,
\begin{equation}
\label{est:error}
\sum_{i=1}^{\lceil T/h\rceil } W_2^2(\mu_{i-1}^h,\mu_{i}^h) \leq 2h(E(\mu_0) - \log
\alpha - 1).
\end{equation}
Hence the right-hand side in~\pref{eq:almostweak} approaches zero as $h\rightarrow0$.
\end{Lemma}

\step4{Convergence}
\begin{Lemma}
\label{lemma:convergence}
There exists a sequence $h_j\to0$  and a uniformly
$W_2$-continuous function $\mu^0:[0,\infty)\rightarrow G$
such that for all $T>0$,
\begin{equation}
\label{conv:W2}
\sup_{t\in[0,T]} W_2(\mu^{h_j}(t), \mu^0(t) ) \rightarrow 0 \quad
\text{as } j\rightarrow \infty.
\end{equation}
\end{Lemma}

\step5{Conclusion}
\begin{Lemma}
\label{lemma:existence}
The limit $\mu^0$ is a weak solution.
\end{Lemma}

We now proceed to prove these lemmas.

\subsection{Proof of Lemma~\ref{lemma:exminpb}}
\label{se:prooflemma:exminpb}

We first state and prove an auxiliary lemma.
\begin{Lemma}
\label{lemma:GMcompact}
For any $\sigma\in G$ and $M>0$ the set
\[
G^M := \left\{\mu\in G: E(\mu)\leq M \text{ and } W_2(\sigma,\mu)\leq M\right\}
\]
is compact with respect to the metric $W_2$.
\end{Lemma}

\begin{proof}
Since the bound on $W_2(\sigma,\mu)$ implies bounds on the support, $G^M$ is tight and therefore precompact with respect to measure convergence; the same bounds on the support also imply that second moments of elements in $G^M$ are uniformly bounded, and therefore $G^M$ is precompact with respect to $W_2$.

We next show that a limit point $\mu$ of $G^M$ has the same structure as $G$ itself, i.e. that $\mu$ can be written as
\begin{equation}
\label{cond:G}
\mu = \rho \Lebesgue + \beta\delta_L + \beta \delta_R,
\end{equation}
for some $\rho\in L^1(\R)$ and $L,R\in \R$, with $\supp\rho\subset [L,R]$. Taking a sequence $\mu^n = (\rho^n, L^n,R^n)\in G^M$, the left and right boundaries converge since by~\pref{decomp_W}
\[
\beta (L^n-L^m)^2 +\beta(R^n-R^m)^2 \leq W_2^2(\mu^n,\mu^m),
\]
by which $(L^n)_n$ and $(R^n)_n$ are Cauchy sequences with limits (say) $L$ and $R$.

As in the proof of Lemma~\ref{lemma:contraction} the bound on $\int \rho^n\log\rho^n$ implies $\sigma(L^1,L^\infty)$-precompactness, so that  there is a subsequence that converges weakly in the $\sigma(L^1,L^\infty)$ topology to a limit $\rho$. This convergence implies $W_2$-convergence of $\rho^n$, since the bounds on $L^n$ and $R^n$ imply a bound on the second moment of $\rho^n$. Combining this with the convergence of the left and right boundaries implies that $\mu$ satisfies~\pref{cond:G}.
The fact that $\supp \rho \in [L,R]$ is a direct consequence of the measure convergence.

Finally, the lower semincontinuity of $E$ (Lemma~\ref{lemma:Elsc}) implies that any limit point $\mu$ of $G^M$ satisfies $E(\mu)\leq M$; therefore $G^M$ is $W_2$-closed.
\end{proof}

By this Lemma the $W_2$-lower semicontinuous functional $I_{h,\sigma}$ has a global minimizer $\mu_0$ in $G$. We now show that $\mu_0$ satisfies~\pref{eq:approxEL1}. We construct a perturbation by the method introduced in~\cite{JordanKinderlehrerOtto98} (see also~\cite{Otto01} or~\cite{Agueh05}). We choose $\psi\in C_c^\infty(\R)$ and
define the flow map $(\Phi_\eps)_{\eps\in\mathbb{R}}$ in
$C^\infty(\mathbb{R})$ by the pointwise differential equation in $\e$,
\[
\left\{
\begin{array}{rcl}
\dfrac{\partial \Phi_\eps}{\partial \eps} &=&\psi \circ \Phi_\eps,\\[2\jot]
\Phi_0 &=& \text{Id}.
\end{array}\right.
\]
Then $\Phi_\eps$ is a $C^1$-diffeomorphism for all
$\e\in \mathbb{R}$, and
\[
\left.\frac{\partial}{\partial \eps} \frac{\partial
\Phi_\eps(x)}{\partial
 x}\right|_{\eps=0} = \frac{\partial}{\partial x} \left.\frac{\partial
 \Phi_\eps(x)}{\partial \eps}\right|_{\eps=0} = \left.\frac{\partial}{\partial
 x} (\psi(\Phi_\eps(x))\right|_{\eps=0} =
 \frac{\partial}{\partial x} \psi(x).
\]
Next we define the probability measure
$\mu_\eps:=(\Phi_\eps)_{\#} \mu_0$. Since $\Phi_\eps$ is monotonic, $\mu_\eps \in
G$.

Let $\mu_\eps=(\rho_\eps,L_\eps,R_\eps)$ and $\sigma =
(\rho,L,R)$. Then $\rho_\eps =
(\Phi_\eps)_{\#}\rho_0$, $L_\eps = \Phi_\eps(L_0)$, and
$R_\eps = \Phi_\eps(R_0)$.
Because $\mu_0$ minimizes $I_{h,\sigma}$ we have
\begin{equation}
\label{eq:ddI} \frac{I_{h,\sigma}(\mu_\eps)-I_{h,\sigma}(\mu_0)}{\eps} \geq
0.
\end{equation}
Following the calculations in any of~\cite{JordanKinderlehrerOtto98,Otto01,Agueh05} we have
\[
\frac12 \frac d{d\e} W_2^2(\rho_\e,\rho)\Bigr|_{\e=0} =
\int(x-y) \psi(x)\, \dd \tilde{\gamma}(x,y),
\]
where $\tilde\gamma$ is the optimal measure in $\Gamma({\rho_0},\rho)$. Combining this with~\pref{decomp_W} we obtain
\begin{eqnarray}
\nonumber \frac{1}{2} \frac{\dd}{\dd\eps} W_2^2(\mu_\eps,\sigma)\Bigr|_{\e=0} &=&
\int(x-y) \psi(x) \dd \tilde{\gamma}(x,y) + \beta (R_0-R)
\left.\frac{\dd}{\dd\eps} \Phi_\eps(R_0)\right|_{\eps=0} + \beta
(L_0 - L) \left.\frac{\dd}{\dd\eps}\Phi_\eps(L_0)\right|_{\eps=0}\\
\nonumber &=& \int(x-y) \psi(x) \dd \tilde{\gamma}(x,y) + \beta
(R_0-R) \psi(R_0) + \beta (L_0 - L) \psi(L_0)\\
\nonumber &=&  \int(x-y) \psi(x) \dd \gamma(x,y).
\end{eqnarray}
Here $\gamma$ is the optimal measure
in $\Gamma(\mu_0, \sigma)$.

Furthermore, again following~\cite{JordanKinderlehrerOtto98} or~\cite{Agueh05}, we have
\begin{equation}
\label{limit:tangent-E}
\frac d{d\e}\int_\mathbb{R} \rho_\eps \log \rho_\eps \,\dd x \Bigr|_{\e=0}
=
-\int_\mathbb{R} \rho_0(x) \psi'(x) \dd x.
\end{equation}
Finally,
\begin{eqnarray}
\nonumber \left. \frac{\dd}{\dd \eps} \alpha R_\eps\right|_{\eps=0}
&=& \alpha
\psi(R_0),\\
\nonumber -\left. \frac{\dd}{\dd \eps} \alpha
L_\eps\right|_{\eps=0}&=& - \alpha \psi(L_0).
\end{eqnarray}
Hence, if we let $\eps$ approach $0$ in (\ref{eq:ddI}), we obtain
\[
\frac{1}{h} \int (x-y) \psi(x)\, \dd \gamma(x,y)
 -
 \int_\mathbb{R} \rho_0(x) \psi'(x)\, \dd x  + \alpha
\psi(R_0) - \alpha \psi(L_0) \geq 0.
\]
Since $\psi$ is arbitrary, actually equality holds. Now take $\zeta\in C_c^\infty(\mathbb{R})$ and set $\psi=\zeta'$. We find
\[
\frac{1}{h} \int (x-y) \zeta' (x) \, \dd \gamma(x,y)
 -
 \int_\mathbb{R} \rho_0(x) \zeta''(x)\, \dd x  + \alpha
\zeta'(R_0) - \alpha \zeta'(L_0) = 0.
\]
We use that
\[
\frac{1}{h}\int_\mathbb{R} \zeta(y) \dd (\mu_0(y) - \sigma(y)) =
\frac{1}{h} \int  (\zeta(x) -\zeta(y)) \dd \gamma(x,y),
\]
and that
\[
\zeta(y) = \zeta(x) + (y-x) \zeta'(x) +\frac{1}{2} (y-x)^2
\zeta''( \theta_{xy}x + (1-\theta_{xy})y ).
\]
where $\theta_{xy}\in[0,1]$, to find~\pref{eq:approxEL1}.

\subsection{Proof of Lemma~\ref{lemma:approxEL}}

This Lemma follows from a simple rearrangement of the discrete time-derivative terms.
Let $\xi \in
C_c^\infty(\mathbb{R}\times\mathbb{R})$, and let $T$ be such that
$\xi(t,x)=0$ for all $t\geq T$ and $x \in \mathbb{R}$. Assume,
without loss of generality, that $T/h \in \mathbb{N}$. By~\pref{eq:approxEL1}, for all $t\in(0,T]$ and $i$ such that $t \in
[t_{i-1},t_i)$,

\begin{multline}
\nonumber \left| \frac{1}{h} \int_\mathbb{R} \xi(t,x) \dd(\mu_i^h(x)
- \mu_{i-1}^h(x)) - \int_\mathbb{R} \rho_i^h(x) \xi_{xx}(t,x) \dd x
+ \alpha \xi_x(t,R_i^h) - \alpha \xi_x(t,L_i^h) \right| \\ \leq
\frac{1}{2h} \| \xi_{xx} \|_\infty W_2^2(\mu_{i-1}^h,\mu_i^h).
\end{multline}
If we integrate this inequality over $t\in(0,T)$ and use the
triangle inequality, we obtain
\begin{multline}
\label{eq:intel} \left| \sum_{i=1}^{T/h}
\int_{t_{i-1}}^{t_i}\left(\frac{1}{h} \int_\mathbb{R} \xi(t,x) \dd
(\mu_i^h(x)-\mu_{i-1}^h(x) )- \int_\mathbb{R}
\rho_i^h(t,x)\xi_{xx}(t,x) \dd x
+ \alpha \xi_x(t,R_i^h)-\alpha \xi_x(t,L^h_i)\right) \dd t \right| \\
\leq \frac{1}{2} \| \xi_{xx} \|_\infty \sum_{i=1}^{T/h}
W_2^2(\mu_{i-1}^h, \mu_i^h).
\end{multline}
Since
\[
\sum_{i=1}^{T/h}  \int_{t_{i-1}}^{t_i}\frac{1}{h}
\int_\mathbb{R}
\xi(t,x) \dd (\mu_i^h(x)-\mu_{i-1}^h(x) ) =  - \int_0^T \int_\mathbb{R}
\partial_t^h \xi(t,x) \dd \mu^h(x)  \dd t - \frac{1}{h} \int_0^h
\int_\mathbb{R} \xi(t,x) \dd \mu_0(x) \dd t,
\]
we deduce~\pref{eq:almostweak} from \pref{eq:intel}.

\subsection{Proof of Lemma~\ref{lemma:est_error}}

For each $i\in\mathbb{N}$, $\mu^h_i$ minimizes $I_{h,\mu^h_{i-1}}$, so that
\[
I_{h,\mu^h_{i-1}}(\mu^h_i)\leq I_{h,\mu^h_{i-1}}(\mu^h_{i-1}) = E(\mu^h_{i-1}),
\]
or
\[
\frac{1}{2h} W^2_2(\mu_{i-1}^h,\mu_i^h) \leq E(\mu^h_{i-1}) -
E(\mu_i^h).
\]
We use a telescoping-sum argument and Lemma
\ref{P:lbintenergy} to see that
\[
\sum_{i=1}^{T/h}W_2^2(\mu_{i-1}^h,\mu_{i}^h) \leq 2h (E(\mu^h_0) -
E(\mu_{T/h}))\leq 2h (E(\mu_0) - \log \alpha - 1).
\]

\subsection{Compactness and proof of Lemma~\ref{lemma:convergence}}

The Arzel\`a-Ascoli theorem is based on a combination of equicontinuity in time with compactness of time slices. In the case at hand, the functions $t\mapsto \mu^h(t)$ are not continuous, but satisfy a weaker condition that is expressed in the following lemma.

\begin{Lemma}
\label{le:timebound} There exists a constant $C$, depending only on
$\rho_0$, $L_0$, $R_0$ and $\alpha$, such that for all $h>0$ and
$t_2\geq t_1\geq0$,
\begin{equation}
\label{ineq:nearly-ec}
W_2\left(\mu^h(t_2),\mu^h(t_1)\right) \leq C \sqrt{t_2-t_1+h}.
\end{equation}
\end{Lemma}
One might think of~\pref{ineq:nearly-ec} as an `equi-near-continuity' condition; for finite $h>0$ jumps are possible, but the maximal size of the jumps converges to zero as $h\to0$.

Since Lemma~\ref{le:timebound} also provides a bound
\[
\sup_{t\in[0,T]} W_2(\mu^h(t),\mu_0) \leq C\sqrt{T+1},
\qquad\text{uniformly in }h\in (0,1],
\]
the set $\bigl\{\,\mu^h(t): h\in (0,1], \ t\in[0,T]\,\bigr\}$ is a subset of $G^M$, with
$M = \max\bigl\{E(\mu_0),C\sqrt{T+1}\bigr\}$, and therefore precompact in $W_2$ by Lemma~\ref{lemma:GMcompact}.
With this compactness and Lemma~\ref{le:timebound} the proof of the Arzel\`a-Ascoli theorem is easily adapted to yield the existence of a sequence $h_j\to 0$ and a limit $\mu^0:[0,\infty)\to G$ such that for any $T>0$,
\[
\sup_{t\in[0,T]} W_2(\mu^{h_j}(t), \mu^0(t) ) \rightarrow 0 \quad
\text{as } j\rightarrow \infty,
\]
thereby proving Lemma~\ref{lemma:convergence}.

\begin{proof}[Proof of Lemma~\ref{le:timebound}]
Define
\[
i_1 := \lfloor t_1/h \rfloor,\quad i_2 := \lfloor t_2/h \rfloor,
\]
so that
\[
i_2 - i_1 \leq \frac{t_2 - t_1}h + 1.
\]
Applying the triangle inequality, the Cauchy-Schwarz inequality, and the bound on the difference in internal energies that
was also used in the previous theorem, we find
\begin{eqnarray}
\nonumber W_2\left(\mu^h(t_2),\mu^h(t_1)\right) &=&
W_2\left(\mu^h_{i_2},\mu^h_{i_1}\right)\\
\nonumber &\leq& \sum_{j=i_1+1}^{i_2}
W_2\left(\mu_j^h,\mu_{j-1}^h\right)\\
\nonumber &\leq&\sqrt{i_2-i_1} \left(\sum_{j=i_1+1}^{i_2}
W_2^2\left(\mu_j^h,\mu_{j-1}^h\right)\right)^{1/2}\\
\nonumber&\leq&
\frac{1}{\sqrt{h}}\sqrt{t_2-t_1+h}\bigl[2h\left(E(\rho^h_{i_1},L^h_{i_1},R^h_{i_1})
-
E(\rho^h_{i_2},L^h_{i_2},R^h_{i_2})\right)\bigr]^{1/2}\\
\nonumber &=& C \sqrt{t_2-t_1+h}.
\end{eqnarray}
with
\[
C := \sqrt 2\, \sqrt{E(\rho_0,L_0,R_0)-\log \alpha + 1}.
\]
\end{proof}

\subsection{Proof of Lemma~\ref{lemma:existence}}

To show that the limit $\mu^0$  is a weak solution, we first note that part~\ref{cond:weaksol:1} of Definition~\ref{de:weak} is satisfied by the definition of $G$ and the $W_2$-continuity of $\mu^0$, and that part~\ref{cond:weaksol:2} follows from the bound
\[
E(\mu^0(t)) \leq \liminf_{j\to\infty} E(\mu^{h_j}(t))
\leq E(\mu_0).
\]
To show~\pref{eq:weakform1}, we first note that from~\pref{conv:W2} and~\pref{decomp_W} follows that for any $T>0$
\begin{alignat*}2
\mu^{h_j}(t)\weakto \mu^0(t) \text{ and }\rho^{h_j}(t) &\weakto \rho^0(t) &\qquad &\text{ as measures, for all $t\in [0,T]$}\\
L^{h_j}, R^{h_j} &\to L^0, R^0 &\qquad &\text{in }L^\infty([0,T]).
\end{alignat*}
With this convergence, and the estimate~\pref{est:error}, we take the limit in~\pref{eq:almostweak} to find~\pref{eq:weakform1}.

\section{Alternative proof}
\label{sec:AGS}

Our proof of Theorem \ref{th:existence} was largely self-consistent. Yet, we see advantages in putting our work in the framework of the book on gradient flows by Ambrosio, Gigli and Savar\'e~\cite{AmbrosioGigliSavare05}, to which we will refer in this section. Because the functional $E$ is lower-semicontinuous (see Lemma \ref{lemma:Elsc}), and $W_2$-bounded subsets of of a sublevel of $E$ are relatively compact (Lemma \ref{lemma:GMcompact}), Assumptions 2.1a and 2.1c from~\cite{AmbrosioGigliSavare05} on the functional are fulfilled, while Property 2.1b is immediate. Since the energy $E$ is geodesically convex,~\cite[Assumption 2.4.5]{AmbrosioGigliSavare05} is fulfilled as well. Consequently, we may conclude that our discretisation scheme yields the so-called curve of maximal slope $\mu:\mathbb{R}^+\to G$ with respect to the strong upper gradient of our energy functional $E$ (see~\cite[Corollary 2.4.11]{AmbrosioGigliSavare05}). Due to the geodesic convexity of $E$, this strong upper gradient in a point $\mu \in G$ equals both the local and the global slope of $E$ in $\mu$, that are defined by (see also~\cite[Definition 1.2.4]{AmbrosioGigliSavare05})
\[
|\partial E|(\mu) := \limsup_{\nu\to\mu} \frac{(E(\mu)-E(\nu))_+}{W_2(\mu,\nu)}, \qquad \varUpsilon_E(\mu) := \sup_{\nu \neq \mu}  \frac{(E(\mu)-E(\nu))_+}{W_2(\mu,\nu)}.
\]
In the following proposition we will compute the strong upper gradient of the internal energy $E$.
\begin{Proposition} 
Let $\mu= (\rho,L,R)\in G$. Then
\[
\varUpsilon_E(\mu) = \begin{cases} \displaystyle \left(\int \frac{(\rho_x)^2}{\rho}+\frac{(\rho(R)-\alpha)^2}{\beta} + \frac{(\rho(L)-\alpha)^2}{\beta}\right)^{1/2}, & \displaystyle \frac{\rho'}{\rho} \in L^2(\rho), \\ 
\infty, & \displaystyle\frac{\rho'}{\rho} \notin L^2(\rho).\end{cases}
\]
\end{Proposition}

\begin{proof} Let $\mu = (\rho,L,R)\in G$. We will first bound $\varUpsilon_E(\mu)$ from below by estimating
\begin{align*}
\varUpsilon_E(\mu) &= \sup_{\nu \neq \mu}  \frac{(E(\mu)-E(\nu))_+}{W_2(\mu,\nu)}\\
&\geq \sup_{T \in C^1_b}  \frac{E(\mu)-E(T_\#\mu)}{W_2(\mu,T_\#\mu)}\\
&\geq \sup_{T \in C^1_b}  \frac{E(\mu)-E(T_\#\mu)}{\sqrt{\int (x-T(x))^2 d\mu(x)}}\\
&= \sup_{T \in C^1_b} \sup_{0<t\leq1} \frac{E(\mu)-E(T_{t\#}\mu)}{\sqrt{\int (x-T_t(x))^2 d\mu(x)}}\\
&= \sup_{T \in C^1_b} \sup_{0<t\leq1} \frac{E(\mu)-E(T_{t\#}\mu)}{t\sqrt{\int (x-T(x))^2 d\mu(x)}},
\end{align*}
where $T_t(x) := (1-t)x+tT(x)$. Furthermore
\[
\sup_{0<t\leq 1} \frac{E(\mu)-E(T_{t\#}\mu)}t \geq \lim_{t\downarrow 0} \frac{E(\mu)-E(T_{t\#}\mu)}{t}.
\]
The limit on the right-hand side equals
\[
\lim_{t\downarrow 0} \frac{E(\mu)-E(T_{t\#}\mu)}t
= \int (T'(x)-1)\rho(x)\,dx - \alpha(T(R) - R) + \alpha(T(L)-L).
\]
Consequently,
\begin{align}
\nonumber \varUpsilon_E(\mu) 
&\geq \sup_{T\in C^1_b} \frac{\int (T'(x)-1)(x)\rho(x)dx - \alpha(T(R) - R) + \alpha(T(L)-L)}{\sqrt{\int (x-T(x))^2 d\mu(x)}}\\
\label{eq:supoverS}&= \sup_{S\in C^1_b} \frac{\int S'(x)\rho(x)dx - \alpha S(R) + \alpha S(L)}{\|S\|_{L^2(\mu)}}.
\end{align}
Suppose this supremum is finite. It follows that the operator $C_b^1 \to \mathbb{R}$ given by 
\[
S \mapsto \int S'(x) \rho(x) dx - \alpha S(R) + \alpha S(L)
\]
can be extended to a bounded operator $L^2(\mu) \to \mathbb{R}$, which in turn can be represented by an $L^2(\mu)$ inner product with a vector. From this it follows that $\rho'/\rho \in L^2(\rho)$. We therefore have proved that if $\rho'/\rho \notin L^2(\rho)$, then $\varUpsilon_E(\mu) = \infty$. 

We now continue under the assumption that the supremum (\ref{eq:supoverS}) is finite, and therefore that $\rho'/\rho \in L^2(\rho)$. Note that therefore the boundary values $\rho(R)$ and $\rho(L)$ are well-defined. We integrate by parts and apply Cauchy-Schwarz to find
\begin{align}
\nonumber \varUpsilon_E(\mu) 
&\geq \sup_{S\in C^1_b} \frac{-\int S(x)\frac{\rho'(x)}{\rho(x)}\, \rho(x)dx - S(R)(\alpha-\rho(R)) - S(L)(\rho(L)-\alpha)}{\|S\|_{L^2(\mu)}}\\
\label{eq:energyslope}&= \left\|\left(\frac {\rho'}\rho,\frac1\beta(\rho(L)-\alpha),\frac1\beta (\alpha-\rho(R))\right) \right\|_{L^2(\mu)}.
\end{align}

For the upper bound, we note that for each $\nu$ and $\varepsilon > 0$ there exist $\nu^\varepsilon$ that have smooth, strictly positive densities and $|E(\nu)-E(\nu^\varepsilon)|<\varepsilon$ and $W_2(\nu,\nu^\varepsilon)< \varepsilon$. Therefore, for each $\nu \in G$,
\[
\frac{(E(\mu) - E(\nu))_+}{W_2(\mu,\nu)} \leq \sup_{\varepsilon>0} \frac{(E(\mu) - E(\nu^\varepsilon))_+}{W_2(\mu, \nu^\varepsilon)}. 
\]
Because of the regularity of the $\nu^\varepsilon$, it holds that the optimal maps $T^{\nu^\varepsilon}$ such that $T^{\nu^\varepsilon}_\#\mu = \nu_\varepsilon$ are in $C^1_b$. We define $T^{\nu^\varepsilon}_t (x):= (1-t)x + T^{\nu^\varepsilon}(x)$. Due to displacement convexity, if $E(\nu) < E(\mu)$,
\begin{align*}
\frac{(E(\mu)-E(\nu))_+}{W_2(\mu,\nu)}  & \leq \sup_{\varepsilon>0} \lim_{t \downarrow 0} \frac{E(\mu)- E((T^{\nu^\varepsilon}_t)_\#\nu^\varepsilon)}{W_2(\mu, (T^{\nu^\varepsilon}_t)_\#\nu^\varepsilon)} \\
&= \sup_{\varepsilon > 0} \frac{\int((T^{\nu^\varepsilon})'(x) - 1)\rho(x) dx - \alpha(T^{\nu^\varepsilon}(R)-R)+\alpha(T^{\nu^\varepsilon}(L)-L)}{\sqrt{\int (T^{\nu^\varepsilon}(x)-x)^2\, d\mu(x)}}.
\end{align*}
Therefore,
\[
\varUpsilon_E(\mu) \leq  \sup_{T\in C^1_b} \frac{\int (T'(x)-1)(x)\rho(x)dx - \alpha(T(R) - R) + \alpha(T(L)-L)}{\sqrt{\int (x-T(x))^2 d\mu(x)}}.
\]
Assuming $\rho'/\rho \in L^2(\rho)$ we can partially integrate and apply Cauchy-Schwarz as before to get
\[
\varUpsilon_E(\mu) \leq \left\|\left(\frac {\rho'}\rho,\frac1\beta(\rho(L)-\alpha),\frac1\beta (\alpha-\rho(R))\right) \right\|_{L^2(\mu)}.
\]
\end{proof}

Having the expression for the strong upper gradient at hand, Theorem 2.3.3 in~\cite{AmbrosioGigliSavare05} provides us with the energy identity
\[
\frac{1}{2} \int_0^T |\mu'|^2(t) dt  + \frac{1}{2} \int_0^T |\partial E|^2(\mu(t)) dt+ E(\mu(T))= E(\mu_0), \text{for all } T>0.
\]
Here, the metric derivative $|\mu'|(t)$ is for almost every $t$ defined by
\[
|\mu'|= \lim_{s \to t} \frac{W_2(\mu(s),\mu(t))}{|s-t|}.
\]
and it is an element of $L^2(0,T)$. At the same time, $|\mu'|=|\partial E|$ at almost every $t$. Therefore, the energy identity gives rigorous sense to the formal result for the dissipation rate (\ref{eq:energydissipation}) we established in the introduction. Finally, the regularizing effect as described in~\cite[Theorem 2.4.15]{AmbrosioGigliSavare05} implies that the solution $\mu(t)$ satisfies $|\partial E|(\mu(t))<\infty$ for all $t>0$. In particular, this means that $\rho_x(t;x)/\rho(t;x) \in L^2(\rho(t))$ for all $t>0$. 

The techniques used in Chapter 2 of \cite{AmbrosioGigliSavare05} to prove existence of minimizing movements are very similar, yet more general than the techniques we have used to prove existence. In Chapter 4 of the book, under the more stringent Assumption 4.0.1, it is shown that even somewhat better results can be obtained. Because in the one-dimensional case, for every measure $\mu$ with finite second moment the function $\nu \mapsto W_2^2(\mu,\nu)$ is $1$-convex along geodesics~\cite[(9.1.7)]{AmbrosioGigliSavare05}, Assumption 4.0.1 follows immediately for the case at hand. This enables us to apply~\cite[Theorem 4.0.4]{AmbrosioGigliSavare05}. In the existence proof, one can use a slightly more general initial condition, that the initial datum is in the closure of the domain of $E$. Moreover, the regularity result becomes stronger with actual bounds on $|\partial E|(\mu)$. 

Another nice result of~\cite[Theorem 4.0.4]{AmbrosioGigliSavare05} is that the solution curve $\mu: \mathbb{R}^+ \rightarrow G$ is the unique solution of the evolution variational inequality (see \cite[(4.0.13)]{AmbrosioGigliSavare05} with $\lambda=0$)
\begin{equation}
\label{eq:EVI}
\frac{1}{2}\frac{d}{dt} W^2_2(\mu(t),\sigma) \leq E(\sigma) - E(\mu(t)), \quad
\text{for a.e. $t>0$, for all } \sigma\in G,
\end{equation}
among all absolutely continuous curves such that $\lim_{t\downarrow0} \mu(t) = \mu_0$. We will now show that such a solution is a weak solution in the sense of Definition \ref{de:weak} as well. Theorem 8.3.1 of~\cite{AmbrosioGigliSavare05} yields the existence of a vector field $v(t,\cdot)$ such that the following formulation of the continuity equation is fulfilled:
\begin{equation}
\label{eq:continuity}
\int_\mathbb{R} \int_\mathbb{R} (\zeta_t(t,x) + v(t,x) \zeta_x(t,x))\,d \mu(t;x) dt = 0, \quad \quad \text{for all } \zeta \in C_c^\infty(\mathbb{R}^+\times \mathbb{R}).
\end{equation}
Expressed in this vector field, the derivative in equation (\ref{eq:EVI}) reads
\[
\frac12\frac{d}{dt} W^2_2(\mu(t),\sigma) = \int_\mathbb{R} v(t,x) (x - T^\sigma(t,x)) d\mu(t),
\]
where $T^\sigma(t,\cdot):\mathbb{R}\rightarrow\mathbb{R}$ is the map that pushes $\mu(t)$ forward to $\sigma$. We take an arbitrary function $\zeta\in C_c^\infty(\mathbb{R}\times\mathbb{R}^+)$ and we introduce the flow maps
\[
\left\{
\begin{array}{rcl}
\dfrac{\partial \Phi_\eps^t}{\partial \eps} &=&\zeta_x(t,\cdot) \circ \Phi_\eps^t,\\[2\jot]
\Phi_0 &=& \text{Id}.
\end{array}\right.
\]
Subsequently, we define $\sigma^t_\eps := {\Phi_\eps^t}_\# \mu_t$, and we substitute this for $\sigma$ in the evolutional variational inequality (\ref{eq:EVI}). By dividing by $\eps$ and letting $\eps\to 0$ (compare Section \ref{se:prooflemma:exminpb}), we derive
\[
- \int_\mathbb{R} v(t,x) \zeta_x(t,x) d \mu(t) = - \int_\mathbb{R} \rho(t,x) \zeta_{xx}(t,x) + \alpha \zeta_x(t,R(t)) - \alpha \zeta_x(t,L(t) ).
\]
Next, we integrate over time, and use the continuity equation (\ref{eq:continuity}) to perform a partial integration on the left hand term. With that we have proved that $\mu(t)$ is a weak solution according to Definition~\ref{de:weak}. 

In Lemma \ref{lemma:EVI} we have already proved that a weak solution in the sense of Definition~\ref{de:weak} is a solution to the evolution variational inequality in distributional sense. It is important to note that we did not assume weak solutions to be absolutely continuous. Consequently, the result of uniqueness was not implied by the uniqueness of solutions to (\ref{eq:EVI}). Combining the results of the paper, however, we can state that there is exactly one weak solution, and that solution satisfies~(\ref{eq:EVI}). 

\section{Acknowledgement}

The authors gratefully acknowledge several helpful discussions with Prof. dr. Giuseppe Savar\'e.

\appendix

\section{A heuristic explanation of the Wasserstein gradient flow}
\label{app:heuristic}

We first recall the structure of a gradient flow in Riemannian geometry.
On a smooth Riemannian manifold $M$ the gradient of a functional $E:M\to\R$ is defined by the condition
\[
g_\rho(\grad E (\rho), s) = E'(\rho)\cdot s\qquad\text{for all }s\in T_\rho M.
\]
Here $T_\rho M$ is the tangent plane at $\rho\in M$, $g_\rho$ the local metric tensor, and $E'(\rho)$ is the Fr\'echet derivative (or differential) of $E$ at $\rho$. The gradient flow of $E$ on $M$ is then the evolution equation given by
\[
\partial _t \rho(t) = -\grad E(\rho(t)),
\]
or equivalently
\begin{equation}
\label{def:abstract_GF}
g_{\rho(t)}(\partial_t\rho(t), s) = -E'(\rho(t))\cdot s\qquad\text{for all }s\in T_{\rho(t)} M.
\end{equation}

Essentially, we wish to apply this abstract concept to the case of the Wasserstein metric. Unfortunately, this is not the metric of a smooth Riemannian manifold. Strictly speaking, the concept and language of gradient flows on Riemannian manifolds are therefore unavailable for this metric (although various ways are known to circumvent this problem~\cite{Otto01,BenamouBrenier00,CarrilloMcCannVillani06,AmbrosioGigliSavare05}). Despite this fact, the setting of Riemannian gradient flows does provide the best insight for deriving formulas and formulating conjectures, and we therefore choose this setting to motivate the results.

\medskip
At any fixed time $t$, we consider the unknown in the free boundary problem to be an element of the set
\[
M := \left\{ (\Omega,\rho): \rho\in L^1(\R^d),
 \ \supp\rho\subset \Omega\subset\R^d, \ \int_\Omega \rho = m \right\},
\]
for some fixed $m>0$. If $t\mapsto (\Omega(t),\rho(t,\cdot))$ is a curve in $M$, then for any $\phi\in C(\R^d)$ we have
\[
\frac d{dt} \int_{\Omega(t)} \phi(x)\rho(t,x)\, dx
 = \int_{\partial\Omega(t)} \phi(x) \rho(t,x) \partial_t\Omega(t,x)\, ds(x)
   + \int_{\Omega(t)} \phi(x)\partial_t \rho(t,x)\, dx.
\]
Here we use the notation $\partial_t\Omega(t,x)$ for the (exterior) normal velocity of $\partial \Omega(t)$ at the point $x$.
At any $(\Omega,\rho)\in M$, we therefore identify the tangents to $M$ to a pair $(v,s)$, where $v:\partial \Omega\to\R$ is the normal velocity of $\partial\Omega$ and $s\in L^1(\Omega)$ represents the change in $\rho$. The conservation of total mass implies that  at a point $(\Omega,\rho)\in M$ all admissible pairs $(v,s)$ should satisfy
\[
\int_{\partial\Omega} \rho(x) v(x)\, ds(x)
   + \int_\Omega s(x)\, dx =0.
\]

On $M$ we define the energy function
\[
E(\Omega,\rho) := \int_\Omega [\rho\log\rho + \alpha],
\]
where $\alpha>0$ is a constant, and as (formal) Riemannian metric tensor we choose
\begin{equation}
\label{def:g}
g_{(\Omega,\rho)} ((v_1,s_1),(v_2,s_2)) :=
  \int_{\Omega} \rho(x) \nabla p_1(x)\nabla p_2(x)\, dx
  + \int_{\partial \Omega} h(\rho(x)) v_1(x)v_2(x)\, ds(x).
\end{equation}
Here $h$ is a positive function that will be related to $f$---see below---and $p_1$ and $p_2$ are related to the tangents $(v_1,s_1)$ and $(v_2,s_2)$ by
\begin{alignat*}3
&s_1 + \div \rho\nabla p_1 = 0,  &\qquad&s_2 + \div \rho\nabla p_2 = 0 &\qquad & \text{in }\Omega\\
&\frac{\partial p_1}{\partial n} = v_1, &\qquad&\frac{\partial p_2}{\partial n} = v_2 & &\text{on }\partial \Omega.
\end{alignat*}

We claim that formally this gradient flow reduces to~\pref{pb:mainNd}. To show this, let $t\mapsto (\Omega(t),\rho(t,\cdot))$ be a path in $M$, satisfying at each time $t$ the gradient-flow equation~\pref{def:abstract_GF}, which becomes
\begin{equation}
\label{def:GF}
g_{(\Omega(t),\rho(t))} ((\partial_t\Omega,\partial_t\rho),(v_2,s_2))
= - E'(\Omega(t),\rho(t))\cdot (v_2,s_2)
\qquad\text{for all } (v_2,s_2)\text{ and }t>0.
\end{equation}
We now choose a $t>0$ fixed. The first term in~\pref{def:GF} is by definition
\[
\int_{\Omega(t)} \rho(t,x)\nabla p_1(t,x)\nabla p_2(x)\, dx
+ \int_{\partial\Omega(t)} h(\rho(t,x))\partial_t \Omega(t,x) v_2(x)\, ds(x),
\]
where
\begin{alignat*}3
&\partial_t\rho(t,\cdot) + \div \rho(t,\cdot)\nabla p_1 = 0,  &\qquad&s_2 + \div \rho\nabla p_2 = 0 &\qquad & \text{in }\Omega(t)\\
&\frac{\partial p_1}{\partial n} = \partial_t\Omega(t,\cdot), &\qquad&\frac{\partial p_2}{\partial n} = v_2 & &\text{on }\partial \Omega(t).
\end{alignat*}
By partial integration these terms can be written as
\begin{equation}
\label{gf:lhs}
\int_{\Omega(t)} \partial_t \rho(t,x)p_2(x)\, dx
  + \int_{\partial\Omega(t)} \rho(t,x)\frac{\partial p_1}{\partial n} (x)p_2(x)\,ds(x)
  + \int_{\partial \Omega(t)}h(\rho(t,x))\partial_t \Omega(t,x)v_2(x)\, ds(x).
\end{equation}
On the other hand, suppressing variables $x$ and $t$ for readability,
\begin{align*}
E'(\Omega,\rho)\cdot (v_2,s_2)
&= \int_{\Omega} (\log\rho+1)s_2
    + \int_{\partial \Omega} (\rho\log\rho+\alpha )v_2\\
&= -\int_{\Omega}(\log\rho+1)\div\rho\nabla p_2
  + \int_{\partial \Omega} (\rho\log\rho+\alpha )v_2\\
&= \int_{\Omega}\nabla\rho\nabla p_2
  - \int_{\partial\Omega} (\log\rho+1)\rho \frac{\partial p_2}{\partial n}
  + \int_{\partial \Omega} (\rho\log\rho+\alpha )v_2\\
&= -\int_{\Omega}p_2\Delta\rho
  + \int_{\partial\Omega} p_2\frac{\partial \rho}{\partial n}
  + \int_{\partial \Omega} (\alpha-\rho)v_2  .
\end{align*}
Combining this with~\pref{gf:lhs} and~\pref{def:GF} we find
\[
0 = \int_\Omega(\partial_t\rho -\Delta\rho)p_2
  + \int_{\partial\Omega} \left(\rho\partial_t\Omega+\frac{\partial\rho}{\partial n}\right)p_2
  + \int_{\partial\Omega} (h(\rho)\partial_t\Omega + \alpha-\rho)v_2.
\]
By making various choices for $p_2$ and $v_2$ it follows that each of the three expressions in parentheses is identically equal to zero. These three expressions correspond to the three equations of~\pref{pb:mainNd}, in the following form:
\begin{alignat*}2
&\partial_t\rho = \Delta \rho &\qquad& \text{in }\Omega(t)\\
&\frac{\partial\rho}{\partial n} = -\rho v_n&&\text{on }\partial \Omega(t)\\
&v_n = \frac{\rho-\alpha}{h(\rho)} &&\text{on }\partial \Omega(t),
\end{alignat*}
where we have reverted back to writing $v_n$ for the normal velocity.
Note that the function $f$ from~\pref{pb:mainNd:kinetic} is forced to have a specific form, $(\rho-\alpha)/h(\rho)$. Since the function $h$ is necessarily positive, this implies that the sign of $f(\rho)$ is determined by this structure: there should be exactly one critical value of $\rho$ at which the sign of $f$ changes, from negative (for smaller $\rho$) to positive.

Also note how the function $f$ is determined by a combination of energy and metric properties: the sign is determined by the \emph{energy} parameter $\alpha$, while the rest of the rate is given by the \emph{metric} parameter $h(\rho)$.

\section{Proof of Lemma~\ref{lemma:char_varphi}}
\begin{proof}
The transport map $\tilde\varphi$ for $W_2(\rho^0,\rho^1)$ is convex on $\R$ and 
achieves equality in~\pref{eq:dualW_2} for $\rho^{0,1}$, i.e.
\[
\tfrac12 W_2^2(\rho^0,\rho^1) = \int_\R (\tfrac12 x^2 - \tilde\varphi(x)) \rho^0(x)\, dx
      + \int_\R (\tfrac12 y^2 - \tilde\varphi^*(y))\rho^1(y)\, dy.
\]
The map $\tilde\varphi$ is not uniquely defined on $\R\setminus\supp\rho^0$, hence we can adapt it slightly. Let $[\tilde{L}_0,\tilde{R}_0]\subset\R$ be the smallest interval such that $\supp \rho_0 \subset [\tilde{L}^0,\tilde{R}^0]$. We define a new convex function $\varphi$ by:
\[
\varphi(x) := \begin{cases}
\tilde\varphi(x) & \text{if }\tilde{L}^0\leq y\leq \tilde{R}^0\\
\tilde\varphi(\tilde{L}^0) + L^1 (x - \tilde{L}^0)& \text{if } x \leq \tilde{L}^0\\
\tilde\varphi(\tilde{R}^0) + R^1 (x - \tilde{R}^0)& \text{if } x \geq \tilde{R}^0.
\end{cases}
\]
Because $\tilde{\varphi}_x$ is uniquely determined $\rho^0$-a.e. and can be interpreted as a transport map,
\[
\partial \varphi(x) \subset [L^1,R^1], \qquad \text{for all } x \in (\tilde{L}^0,\tilde{R}^0).
\]
Therefore, $\varphi$ is indeed convex, with conjugate $\varphi^*$ equal to
\[
\varphi^*(y) := \begin{cases}
\tilde\varphi^*(y) & \text{if }L^1\leq y \leq R^1\\
\infty & \text{otherwise}.
\end{cases}
\]
For the second derivative of $\varphi^*$, which exists almost everywhere on $(L^1,R^1)$, we have
\[
\varphi^*_{yy}(y) = \tilde\varphi^*_{yy}(y)
= \frac{\rho^1(y)}{\rho^0(\tilde\varphi^*_y(y))} > 0
\qquad\text{for almost all }L^1<y< R^1,
\]
and therefore the function $\varphi^*$ is strictly convex on $[L^1,R^1]$. It follows that $\varphi$ is essentially smooth~\cite[Theorem~26.3]{Rockafellar72}, implying that $\varphi$ is differentiable on $\R$. Since $\varphi$ is a differentiable convex function with bounded derivative, $\varphi_{xx}\in L^1(\R)$, and $\varphi\in W^{2,1}_{\mathrm{loc}}(\R)$. As $\varphi_x(L_0) = \varphi_x(\tilde{L}^0) = L^1$ and $\varphi_x(R^0) = \varphi_x(\tilde R^0) = R^1$, we find $L^0L^1 = \varphi(L^0)+\varphi^*(L^1)$ and $R^0R^1 = \varphi(R^0)+\varphi^*(R^1)$.

Finally we show~\pref{eq:W2-equality}. First note that
\begin{align*}
(L^0-L^1)^2 &= (L^0)^2 - 2L^0L^1 + (L^1)^2 \\
  &=   (L^0)^2 - 2\varphi(L^0) - 2\varphi^*(L^1) + (L^1)^2
\end{align*}
We then have by monotonicity
\begin{align*}
\tfrac12 W_2^2(\mu^0,\mu^1) &= \frac\beta2(L^0-L^1)^2 + \tfrac12 W_2^2(\rho^0,\rho^1)
  + \frac\beta2 (R^0-R^1)^2\\
  &=\frac\beta2(L^0-L^1)^2+\int_\R (\tfrac12 x^2 - \tilde\varphi(x)) \rho^0(x)\, dx
      + \int_\R (\tfrac12 y^2 - \tilde\varphi^*(y))\rho^1(y)\, dy
  + \frac\beta2 (R^0-R^1)^2\\
  &= \frac\beta2\left[(L^0)^2 - 2\varphi(L^0) - 2\varphi^*(L^1) + (L^1)^2\right] + \\
  &\qquad {}+
    \int_\R (\tfrac12 x^2 -\varphi(x)) \rho^0(x)\, dx
      + \int_\R (\tfrac12 y^2 -\varphi^*(y))\rho^1(y)\, dy\\
  &\qquad {}+ \frac\beta2 \left[(R^0)^2 - 2\varphi(R^0)-2\varphi^*(R^1)+(R^1)^2\right]\\
  &= \int(\tfrac12 x^2 - \varphi(x))\, d\mu^0(x)
  + \int (\tfrac12 y^2-\varphi^*(y))\, d\mu^1(y).
\end{align*}
\end{proof}

\bibliographystyle{plain}
\bibliography{PPrefs}

\end{document}